\documentclass[12pt,english]{article}
\usepackage{tgtermes}

\usepackage[T1]{fontenc}
\usepackage[latin9]{inputenc}
\usepackage[a4paper]{geometry}
\usepackage{array}
\usepackage{float}
\usepackage{multirow}
\usepackage{amsbsy}
\usepackage{amstext}
\usepackage{amssymb}
\usepackage{graphicx}
\usepackage[authoryear]{natbib}

\makeatletter

\providecommand{\tabularnewline}{\\}
\floatstyle{ruled}
\newfloat{algorithm}{tbp}{loa}
\providecommand{\algorithmname}{Algorithm}
\floatname{algorithm}{\protect\algorithmname}

\makeatother

\usepackage{babel}
\begin{document}

\title{Bayesian Inference of Local Projections\linebreak{}
 with Roughness Penalty Priors}

\author{Masahiro Tanaka\thanks{Graduate School of Economics, Waseda University. Address: 1-6-1, Nishi-Waseda,
Shinjuku-ku, Tokyo 169-8050 Japan, Email: gspddlnit45@toki.waseda.jp.}}

\date{July 7, 2019}
\maketitle
\begin{abstract}
A local projection is a statistical framework that accounts for the
relationship between an exogenous variable and an endogenous variable,
measured at different time points. Local projections are often applied
in impulse response analyses and direct forecasting. While local projections
are becoming increasingly popular because of their robustness to misspecification
and their flexibility, they are less statistically efficient than
standard methods, such as vector autoregression. In this study, we
seek to improve the statistical efficiency of local projections by
developing a fully Bayesian approach that can be used to estimate
local projections using roughness penalty priors. By incorporating
such prior-induced smoothness, we can use information contained in
successive observations to enhance the statistical efficiency of an
inference. We apply the proposed approach to an analysis of monetary
policy in the United States, showing that the roughness penalty priors
successfully estimate the impulse response functions and improve the
predictive accuracy of local projections.

\bigskip{}



\newpage{}
\end{abstract}

\section{Introduction}

Local projections introduced by \citet{Jorda2005} provide a statistical
framework that accounts for the relationship between an exogenous
variable and an endogenous variable, measured at different time points.
Typical applications of local projections include impulse response
analyses and direct (non-iterative) forecasting \citep{Stock2007}.
A local projection has several advantages over standard methods, such
as vector autoregression (VAR). First, it does not impose a strong
assumption on the data-generating process, making it robust to misspecification.
Second, it can easily deal with asymmetric and/or state-dependent
impulse responses (e.g., \citealp{Riera-Crichton2015,Auerbach2013,Ramey2018}).
On the other hand, local projections have several disadvantages. First,
when using a local projection, the exogenous variable must be identified
beforehand. Second, a local projection is statistically less efficient
than other methods, and typically obtains a wiggly impulse response
function, (e.g., \citealp{Ramey2016a}). In an impulse response analysis,
the shape of an estimated impulse response function is of concern.
Therefore, if an estimated impulse response function is wiggly and
has wide confidence/credible intervals, it is difficult to interpret
the result, and one might wrongly reject or accept a hypothesis .
In this study, we address the second disadvantage of local projections. 

In order to improve the statistical efficiency, we develop a fully
Bayesian approach that can be used to estimate local projections using
roughness penalty priors as well as B-spline basis expansions.\footnote{See, e.g., \citet{Geweke2005} for a general introduction to Bayesian
analysis.} The proposed priors, which are adapted from Bayesian splines \citep{Lang2004},
are generated from an intrinsic Gaussian Markov random field; that
is, they induce random-walk behavior on a sequence of parameters.
By incorporating such prior-induced smoothness, we can use information
contained in successive observations to enhance the statistical efficiency
of an inference. We compare the proposed approach with the existing
approaches through a series of Monte Carlo experiments. The proposed
approach is applied to an analysis of monetary policy shocks in the
United States to show how the roughness penalty priors successively
smooth impulse responses and improve statistical efficiency in terms
of predictive accuracy. Furthermore, we show that such improvements
are almost entirely attributable to the roughness penalty priors and
not to the B-spline expansions. 

There are three strands of studies related to this work. For the first,
\citet{Barnichon2019a} approximate a moving average representation
of a time series using values from Gaussian basis functions. Their
approximation is simpler, but much coarser than ours. As a result,
their estimated impulse responses may be excessively smoothed and
vulnerable to model misspecification. For the second, to smooth an
impulse response estimate, \citet{Miranda-Agrippino2017} penalize
the estimate based on deviations from an impulse response derived
from an estimated VAR. However, their approach seems not to work well
in cases with asymmetric and/or state-dependent impulse responses.
Furthermore, their approach uses the same dataset twice. This shortcoming
can be resolved if a time series is long enough to be split into training
and estimation samples, but this is not the general situation in macroeconomic
studies. In contrast, our approach does not require a reference model,
thus it is free from these problems. 

For the third, the most relevant studies are those of \citet{Barnichon2019}
and \citet{El-Shagi2019}, who develop frequentist methods using roughness
penalties. Although our approach can be regarded as a Bayesian counterpart
to theirs, it confers four additional benefits. First, our approach
is more flexible than \citeauthor{Barnichon2019}'s (2019) approach:
they allow a single parameter to control the smoothness of all parameter
sequences, whereas we can assign different smoothing parameters to
individual sequences. Second, our Bayesian approach can evaluate credible
intervals in a consistent and straightforward manner, while the frequentist
approaches cannot provide a theoretically grounded confidence interval.
Third, in our approach, smoothing parameters are inferred from priors,
implying that we can systematically consider uncertainty in the smoothness
of an impulse response. In contrast, the frequentist approach prefixes
smoothing parameters; \citet{Barnichon2019} choose a smoothing parameter
via cross-validation, while \citet{El-Shagi2019} determines smoothing
parameters on the basis of some information criteria. Fourth, our
approach has better finite-sample performance than \citeauthor{El-Shagi2019}'s
(2019) approach, as shown in Section 5.

The rest of the paper is organized as follows. Section 2 introduces
the model, the priors and the posterior simulation. Section 3 conducts
a set of Monte Carlo experiments and reports the result. Section 4
demonstrates our approach in an analysis of the macroeconomic effects
of monetary policy shocks in the United States. Section 5 compares
the proposed approach with the existing frequentist approaches. Section
6 concludes this paper.

\section{Proposed Approach}

We consider two classes of local projections: those with and those
without B-spline expansions. 

\subsection{Local Projection without B-spline expansions}

\subsubsection{Model}

We begin by describing a local projection \citep{Jorda2005}. While
we consider only time series data, an extension to panel data is straightforward.
A model for an individual observation is given by

\[
y_{t+h}=\beta_{\left(h\right)}z_{t}+\alpha_{\left(h\right)}+\sum_{j=1}^{J-2}\gamma_{j,\left(h\right)}w_{j,t}+u_{\left(h\right),t+h},\;h\in\mathcal{H},\;t=1,...,T,
\]
where $\mathcal{H}=\left\{ h_{1},...,h_{H}\right\} \subseteq\mathbb{N}^{H}$
is a set of projection points such that $h_{1}<h_{2}<\cdots h_{H}$,
$y_{t+h}$ is an endogenous variable observed at period $t+h$, $\alpha_{\left(h\right)}$
is an intercept, $z_{t}$ is an exogenous variable observed at period
$t$, $w_{1,t},...,w_{J-2,t}$ are covariates, which may include lags
of the endogenous and exogenous variables, $\beta_{\left(h\right)}$
and $\gamma_{j,\left(h\right)}$ are unknown coefficients, and $u_{\left(h\right),t+h}$
is a residual. The model allows asymmetric and/or state-dependent
impulse responses, as in \citet{Riera-Crichton2015}, \citet{Auerbach2013},
and \citet{Ramey2018}. The definition of $y_{t+h}$ and $z_{t}$
depends on whether the model is used for an impulse response analysis
or forecasting. For the former task, $y_{t+h}$ denotes a response
observed $h$ periods after shock $z_{t}$ occurs at $t$; for the
latter, $z_{t}$ is one of several predictors observed at $t$, and
$y_{t+h}$ is an $h$-period-ahead target observation. In what follows,
we focus on impulse response analysis.

In an impulse response analysis, we seek to infer a smooth function
$f_{z}\left(h\right)$ that represents an impulse response of $y$
to $z$, namely, $f_{z}\left(h\right)=\partial y_{t+h}/\partial z_{t}$.
Here, we allow a sequence $\left\{ \beta_{\left(h_{1}\right)},...,\beta_{\left(h_{H}\right)}\right\} $
to represent an impulse response of $y$ to $z$, namely, $\beta_{\left(h\right)}=\partial y_{t+h}/\partial z_{t}$.
The model can be represented as 
\[
y_{\left(h\right),t+h}=\boldsymbol{x}_{t}^{\top}\boldsymbol{\theta}_{\left(h\right)}+u_{\left(h\right),t+h},\quad t=1,..,T;\;h\in\mathcal{H},
\]
\[
\boldsymbol{x}_{t}=\left(z_{t},1,w_{1,t},...,w_{J-2}\right)^{\top},
\]
\[
\boldsymbol{\theta}_{\left(h\right)}=\left(\beta_{\left(h\right)},\alpha_{\left(h\right)},\gamma_{\left(h\right),1},...,\gamma_{\left(h\right),J-2}\right)^{\top},
\]
where $\boldsymbol{x}_{t}$ is a vector of regressors and $\boldsymbol{\theta}_{\left(h\right)}$
is a vector of corresponding parameters. For notational convenience,
we reindex the coefficient vector as $\boldsymbol{\theta}_{\left(h\right)}=\left(\theta_{\left(h\right),1},...,\theta_{\left(h\right),J}\right)^{\top}$.
Stacking these over the projection dimension yields a representation
resembling a seemingly unrelated regression (SUR): for $t=1,...,T$,
\[
\boldsymbol{y}_{t}=\left(\boldsymbol{I}_{H}\otimes\boldsymbol{x}_{t}^{\top}\right)\boldsymbol{\theta}+\boldsymbol{u}_{t},
\]
\[
\boldsymbol{y}_{t}=\left(y_{\left(h_{1}\right),t+h_{1}},...,y_{\left(h_{H}\right),t+h_{H}}\right)^{\top},\quad\boldsymbol{u}_{t}=\left(u_{\left(h_{1}\right),t+h_{1}},...,u_{\left(h_{H}\right),t+h_{H}}\right)^{\top},
\]
\[
\boldsymbol{\theta}=\left(\boldsymbol{\theta}_{\left(h_{1}\right)}^{\top},...,\boldsymbol{\theta}_{\left(h_{H}\right)}^{\top}\right)^{\top},
\]
where $\otimes$ denotes the Kronecker product. 

Rearranging the above representation, we express the model in matrix
notation as
\begin{equation}
\boldsymbol{y}=\left(\boldsymbol{I}_{H}\otimes\boldsymbol{X}\right)\boldsymbol{\theta}+\boldsymbol{u},\quad\boldsymbol{u}\sim\mathcal{N}\left(\boldsymbol{0}_{HT},\;\boldsymbol{\Sigma}\otimes\boldsymbol{I}_{T}\right)\label{eq: stacked model}
\end{equation}
\[
\boldsymbol{y}=\left(\boldsymbol{y}_{1}^{\top},...,\boldsymbol{y}_{T}^{\top}\right)^{\top},\quad\boldsymbol{X}=\left(\boldsymbol{x}_{1},...,\boldsymbol{x}_{T}\right)^{\top},\quad\boldsymbol{u}=\left(\boldsymbol{u}_{1}^{\top},...,\boldsymbol{u}_{T}^{\top}\right)^{\top}.
\]
where $\boldsymbol{\Sigma}$ is a covariance matrix, and $\mathcal{N}\left(\boldsymbol{d},\boldsymbol{B}\right)$
denotes a multivariate normal distribution with mean $\boldsymbol{d}$
and covariance $\boldsymbol{B}$. Letting $\mathcal{D}$ denote the
data, the likelihood takes a standard form:
\begin{eqnarray*}
p\left(\mathcal{D}|\boldsymbol{\theta},\boldsymbol{\Sigma}\right) & = & \left(2\pi\right)^{-\frac{HT}{2}}\left|\boldsymbol{\Sigma}\right|^{-\frac{T}{2}}\exp\left[-\frac{1}{2}\tilde{\boldsymbol{u}}^{\top}\left(\boldsymbol{\Sigma}^{-1}\otimes\boldsymbol{I}_{T}\right)\tilde{\boldsymbol{u}}\right]\\
 & = & \left(2\pi\right)^{-\frac{HT}{2}}\left|\boldsymbol{\Sigma}\right|^{-\frac{T}{2}}\exp\left[-\frac{1}{2}\textrm{tr}\left(\tilde{\boldsymbol{U}}^{\top}\tilde{\boldsymbol{U}}\boldsymbol{\Sigma}^{-1}\right)\right],
\end{eqnarray*}
\[
\tilde{\boldsymbol{u}}=\left(\tilde{\boldsymbol{u}}_{\left(h_{1}\right)}^{\top},...,\tilde{\boldsymbol{u}}_{\left(h_{H}\right)}^{\top}\right)^{\top}=\boldsymbol{y}-\left(\boldsymbol{I}_{H}\otimes\boldsymbol{X}\right)\boldsymbol{\theta},
\]
where $\tilde{\boldsymbol{U}}=\left(\tilde{\boldsymbol{u}}_{\left(h_{1}\right)},...,\tilde{\boldsymbol{u}}_{\left(h_{H}\right)}\right)$
is a matrix composed of the realized residuals.

\subsubsection{Bayesian Inference}

This section first discusses priors on the subsets of $\boldsymbol{\theta}$
and then assembles them into a prior on $\boldsymbol{\theta}$, followed
by a description of the priors on the other parameters. Lastly, the
posterior simulation method is discussed.

We introduce a class of roughness penalty priors for $\boldsymbol{\theta}_{j}$,
$j=1,...,J$. Our prior construction is motivated by \citet{Lang2004}.
The prior induces an $r$th-order random-walk behavior on a sequence
of parameters $\theta_{\left(h_{1}\right),j},...,\theta_{\left(h_{H}\right),j}$.
When $r=2$, the relationship between $\theta_{\left(h_{i}\right),j}$
and successive parameters is represented by 
\[
\theta_{\left(h_{i}\right),j}=2\theta_{\left(h_{i-1}\right),j}-\theta_{\left(h_{i-2}\right),j}+\epsilon_{i},\quad\epsilon_{i},\sim\mathcal{N}\left(0,\;\tau_{j}^{-1}\lambda_{\left(h_{i}\right),j}^{-1}\right),
\]
for $i=r+1,...,H$, where $\tau_{j}$ and $\lambda_{\left(h_{i}\right),j}$
are global and local smoothing parameters, respectively. Controlling
local smoothness is potentially beneficial, because impulse response
functions often have both strongly bent and smooth areas: for example,
fast-growing responses immediately after an occurrence of shock and
virtually flat responses after convergence to a long-run equilibrium.
In some applications, without the adaptation for local smoothness,
an estimated impulse response might be oversmoothed in some areas
and undersmoothed in others. A prior on $\boldsymbol{\theta}_{j}$
is an improper normal prior generated by an intrinsic Gaussian Markov
random field \citep{Rue2005}, and the smoothing parameters are inferred
from gamma priors, unlike in existing approaches such as \citet{Miranda-Agrippino2017,Barnichon2019a}.
The hierarchy of the prior takes the form
\begin{eqnarray*}
p\left(\boldsymbol{\theta}_{j}|\tau_{j},\boldsymbol{\Lambda}_{j}\right) & \propto & \exp\left[-\frac{\tau_{j}}{2}\sum_{i=r+1}^{H}\lambda_{\left(h_{i}\right),j}\left(\Delta^{r}\theta_{\left(h_{i}\right),j}\right)^{2}\right]\\
 & = & \exp\left(-\frac{\tau_{j}}{2}\boldsymbol{\theta}_{j}^{\top}\boldsymbol{D}^{\top}\boldsymbol{\Lambda}_{j}\boldsymbol{D}\boldsymbol{\theta}_{j}\right)\\
 & = & \exp\left(-\frac{\tau_{j}}{2}\boldsymbol{\theta}_{j}^{\top}\boldsymbol{Q}_{j}\boldsymbol{\theta}_{j}\right),
\end{eqnarray*}
\[
\lambda_{\left(h_{r+1}\right),j}=1\;\textrm{and}\;\lambda_{\left(h_{i}\right),j}\sim\mathcal{G}\left(\eta_{1},\eta_{2}\right),\;i=r+2,...,H,
\]
\[
\tau_{j}\sim\mathcal{G}\left(\nu_{1},\nu_{2}\right),
\]
where $\boldsymbol{\Lambda}_{j}=\textrm{diag}\left(\lambda_{\left(h_{r+1}\right),j},...,\lambda_{\left(h_{H}\right),j}\right)$,
$\boldsymbol{D}$ is an $\left(H-r\right)$-by-$H$ difference matrix
of order $r$, $\eta_{1}$, $\eta_{2}$, $\nu_{1}$, and $\nu_{2}$
are pre-fixed hyperparameters, $\mathcal{G}\left(a,b\right)$ denotes
a gamma distribution with shape $a$ and rate $b$ (and, thus, with
mean $a/b$ and variance $a/b^{2}$), and $\Delta^{r}$ denotes the
$r$th-order difference operator. By assembling the priors on the
subsets of $\boldsymbol{\theta}$, the prior density for $\boldsymbol{\theta}$
conditional on $\boldsymbol{\tau}=\left\{ \tau_{1},...,\tau_{J}\right\} $
and $\boldsymbol{\Lambda}=\left\{ \boldsymbol{\Lambda}_{1},...,\boldsymbol{\Lambda}_{J}\right\} $
is represented as
\[
p\left(\boldsymbol{\theta}|\boldsymbol{\tau},\boldsymbol{\Lambda}\right)\propto\exp\left(-\frac{1}{2}\boldsymbol{\theta}^{\top}\boldsymbol{Q}\boldsymbol{\theta}\right),
\]
\begin{equation}
\boldsymbol{Q}=\sum_{j=1}^{J}\left(\left(\tau_{j}\boldsymbol{D}^{\top}\boldsymbol{\Lambda}_{j}\boldsymbol{D}\right)\otimes\boldsymbol{E}_{j}\right),\label{eq: prior prec}
\end{equation}
where $\boldsymbol{E}_{j}$ is a $J$-by-$J$ zero matrix in which
the $j$th diagonal element is replaced by one. In what follows, the
above prior is referred to as an adaptive roughness penalty (A-RP)
prior. As a special case, the same prior with all local smoothing
parameters set to one is called a non-adaptive roughness penalty (N-RP)
prior. For the N-RP prior, (\ref{eq: prior prec}) can be rewritten
as
\[
\boldsymbol{Q}=\boldsymbol{D}^{\top}\boldsymbol{D}\otimes\textrm{diag}\left(\tau_{1},...,\tau_{J}\right).
\]

Choosing a prior of the covariance matrix $\boldsymbol{\Sigma}$ is
non-trivial. Because of the strong correlations between the residuals,
$\boldsymbol{\Sigma}$ tends to be close to a matrix of ones and almost
singular. If the Jeffreys prior, a popular non-informative prior for
covariance matrices, is employed, a posterior simulation easily crashes
due to the singularity of the gram matrix of the realized residuals.\footnote{Bayesian inference using the Jeffreys prior for $\boldsymbol{\Sigma}$
almost always fails for the synthetic and real data used in the subsequent
section.} Therefore, prior-induced shrinkage is necessary to complete a posterior
simulation. On the other hand, as \citet{Alvarez2014} argue, an inverse
Wishart prior, another popular choice, can be unintentionally, significantly
informative, resulting in significant biases. For these reasons, we
use a hierarchical inverse Wishart (HIW) prior for $\boldsymbol{\Sigma}$
\citep{Huang2013}:

\[
\boldsymbol{\Sigma}|\boldsymbol{\Phi}\sim\mathcal{IW}\left(2\zeta\boldsymbol{\Phi},\;\zeta+H-1\right),
\]
\[
\Phi=\textrm{diag}\left(\phi_{\left(h_{1}\right)},...,\phi_{\left(h_{H}\right)}\right),
\]
\[
\phi_{\left(h_{i}\right)}\sim\mathcal{G}\left(\frac{1}{2},\upsilon\right),\quad i=1,...,H,
\]
where $\phi_{\left(h_{i}\right)}$ is a hyperparameter to be inferred,
$\zeta$ and $\upsilon$ are prefixed hyperparameters, and $\mathcal{IW}\left(\boldsymbol{A},b\right)$
is an inverse Wishart distribution with scale matrix $\boldsymbol{A}$
and degrees of freedom $b$. This prior distribution is seen as a
scale mixture of inverse Wishart distributions, and is more robust
than an inverse Wishart prior. We conducted a simulation study that
compares an inverse Wishart prior and the HIW prior and show that
the HIW prior has better finite sample performance than an inverse
Wishart prior. See Section A.1 in the ``Online Appendix'' for details. 

We can induce this prior arbitrarily non-informative by setting $\upsilon$
to a very small value, but \citeauthor{Huang2013}'s (2013) recommendation
$\upsilon=10^{-10}$ (in our notation) is too flat to complete the
posterior simulation in this paper. Our default choice in this paper
is $\upsilon=0.01$. Although there is no general procedure to find
a sufficiently small value of $\upsilon$, the results in the subsequent
sections are not sensitive to $\upsilon$ as long as it is chosen
from a fairly large range $\left[10^{-4},10^{-1}\right]$ (see Section
A.2 in the ``Online Appendix''). 

As demonstrated in Section A.2 in the Appendix, the proposed approach
is not very sensitive to the choice of the hyperparameters. However,
as the priors used in this paper are not scale-invariant, a user of
the proposed approach is strongly encouraged to conduct a prior sensitivity
check.

A posterior simulation is conducted using the Markov chain Monte Carlo
(MCMC) algorithm. Because all of the conditional posterior densities
are standard, we can construct a block Gibbs sampler. Each sampling
block is specified as follows. 

\paragraph{Sampling $\boldsymbol{\theta}$}

The conditional posterior density of $\boldsymbol{\theta}$ is given
by the multivariate normal distribution:
\[
\boldsymbol{\theta}|-\sim\mathcal{N}\left(\boldsymbol{m},\boldsymbol{P}^{-1}\right),
\]
\begin{eqnarray}
\boldsymbol{m} & = & \boldsymbol{P}^{-1}\left(\boldsymbol{\Sigma}^{-1}\otimes\boldsymbol{X}^{\top}\right)\boldsymbol{y},\label{eq: cond_mean_theta}\\
\boldsymbol{P} & = & \boldsymbol{\Sigma}^{-1}\otimes\boldsymbol{X}^{\top}\boldsymbol{X}+\boldsymbol{Q}.\label{eq: cond_prec_theta}
\end{eqnarray}
This block is computationally demanding, with two bottlenecks. The
first is concerned with calculation of the prior precision matrix,
which involves repeated high-dimensional matrix multiplications, Eq.
(\ref{eq: prior prec}). The computational cost declines significantly
by treating $\boldsymbol{Q}$ as a sparse matrix. The second bottleneck
is the inversion of $\boldsymbol{P}$. For speed and numerical stability,
we apply the algorithm described in Section 2 of \citet{Rue2001}
(see Algorithm 1), which exploits a banded structure of $\boldsymbol{P}$;
inverting a lower-triangular Cholesky root of $\boldsymbol{P}$, (denoted
by $\boldsymbol{L}$), is faster and more numerically stable than
inverting $\boldsymbol{P}$ itself. 

\begin{algorithm}
\caption{Sampling $\boldsymbol{\theta}$ (Rue 2001) }

\[
\boldsymbol{\theta}\sim\mathcal{N}\left(\boldsymbol{m},\boldsymbol{P}^{-1}\right),
\]
\[
\boldsymbol{m}=\boldsymbol{P}^{-1}\left(\boldsymbol{\Sigma}^{-1}\otimes\boldsymbol{X}^{\top}\right)\boldsymbol{y},\quad\boldsymbol{P}=\boldsymbol{\Sigma}^{-1}\otimes\boldsymbol{X}^{\top}\boldsymbol{X}+\boldsymbol{Q}=\boldsymbol{L}\boldsymbol{L}^{\top}.
\]

\medskip{}

Step 1. Sample $\boldsymbol{a}\sim\mathcal{N}\left(\boldsymbol{0}_{HJ},\boldsymbol{I}_{HJ}\right)$.

Step 2. Solve $\boldsymbol{L}^{\top}\boldsymbol{b}=\boldsymbol{a}$
to obtain $\boldsymbol{b}$.

Step 3. Solve $\boldsymbol{L}\boldsymbol{c}=\left(\boldsymbol{\Sigma}^{-1}\otimes\boldsymbol{X}^{\top}\right)\boldsymbol{y}$
to obtain $\boldsymbol{c}$. 

Step 4. Solve $\boldsymbol{L}^{\top}\boldsymbol{m}=\boldsymbol{c}$
to obtain $\boldsymbol{m}$.

Step 5. Set $\boldsymbol{\theta}=\boldsymbol{b}+\boldsymbol{m}$.
\end{algorithm}

\paragraph{Sampling $\boldsymbol{\tau}$ and $\boldsymbol{\Lambda}$}

The conditional posteriors of the smoothing parameters for $\boldsymbol{\theta}_{j}$,
$j=1,...,J$, are specified as the following gamma distributions:
for $j=1,...,J$,
\[
\tau_{j}|-\sim\mathcal{G}\left(\nu_{1}+\frac{1}{2}\textrm{rank}\left(\boldsymbol{D}^{\top}\boldsymbol{D}\right),\;\nu_{2}+\frac{1}{2}\boldsymbol{\theta}_{j}^{\top}\boldsymbol{D}^{\top}\boldsymbol{\Lambda}_{j}\boldsymbol{D}\boldsymbol{\theta}_{j}\right),
\]
\[
\lambda_{\left(h_{i}\right),j}|-\sim\mathcal{G}\left(\eta_{1}+\frac{1}{2},\;\eta_{2}+\frac{\tau_{j}}{2}\left(\Delta^{r}\theta_{\left(h_{i}\right),j}\right)^{2}\right),\;i=r+2,...,H.
\]

\paragraph{Sampling $\boldsymbol{\Sigma}$ and $\boldsymbol{\Phi}$}

The conditional posteriors of $\boldsymbol{\Sigma}$ and $\boldsymbol{\Phi}$
are 

\[
\boldsymbol{\Sigma}|-\sim\mathcal{IW}\left(2\zeta\boldsymbol{\Phi}+\tilde{\boldsymbol{U}}^{\top}\tilde{\boldsymbol{U}},\;\zeta+H-1+T\right),
\]
\[
\phi_{\left(h_{i}\right)}|-\sim\mathcal{G}\left(\frac{\zeta+T}{2},\;\upsilon+\zeta\left(\boldsymbol{\Sigma}^{-1}\right)_{i,i}\right),\quad i=1,...,H,
\]
where $\left(\boldsymbol{\Sigma}^{-1}\right)_{i,i^{\prime}}$ denotes
the $\left(i,i^{\prime}\right)$-element of $\boldsymbol{\Sigma}^{-1}$.

\subsection{Local projection with B-spline expansions}

We consider a local projection with B-spline expansions as an additional
smoothing device. We intend to approximate an impulse response function
$f_{z}\left(h\right)$ using a B-spline basis function expansion over
a projection horizon\footnote{See, for example, \citet{DeBoor1978,Eilers1996} for a detailed description
of B-splines.}
\[
f_{z}\left(h\right)=\beta_{\left(h\right)}\approx\sum_{k=1}^{K}b_{k}\varphi_{k}\left(h\right)=\boldsymbol{b}^{\top}\boldsymbol{\varphi}\left(h\right),
\]
where $K$ is a number of knots, $\boldsymbol{b}=\left(b_{1},...,b_{K}\right)^{\top}$
is a vector of coefficients, and $\boldsymbol{\varphi}\left(h\right)=\left(\varphi_{1}\left(h\right),...,\varphi_{K}\left(h\right)\right)^{\top}$
is a vector of B-spline basis functions. We define the approximations
of the other coefficients in a similar fashion. Given the approximation,
the model is represented as 
\begin{eqnarray*}
y_{t+h} & \approx & \sum_{k=1}^{K}a_{k}\varphi_{k}\left(h\right)+\sum_{k=1}^{K}b_{k}\varphi_{k}\left(h\right)z_{t}+\sum_{j=1}^{J-2}\sum_{k=1}^{K}c_{j,k}\varphi_{k}\left(h\right)w_{j,t}+u_{\left(h\right),t+h}\\
 & = & \boldsymbol{a}^{\top}\boldsymbol{\varphi}\left(h\right)+\boldsymbol{b}^{\top}\boldsymbol{\varphi}\left(h\right)z_{t}+\sum_{j=1}^{J-2}\boldsymbol{c}_{j}^{\top}\boldsymbol{\varphi}\left(h\right)w_{j,t}+u_{\left(h\right),t+h}\\
 & = & \boldsymbol{\vartheta}^{\top}\left(\breve{\boldsymbol{x}}_{t}\otimes\boldsymbol{\varphi}\left(h\right)\right)+u_{\left(h\right),t+h},
\end{eqnarray*}
where $\breve{\boldsymbol{x}}_{t}=\left(z_{t},1,w_{1,t},...,w_{J-2,t}\right)^{\top}$
is a vector of regressors, and $\boldsymbol{\vartheta}=\left(\boldsymbol{b}^{\top},\boldsymbol{a}^{\top},\boldsymbol{c}_{1}^{\top},...,\boldsymbol{c}_{J-2}^{\top}\right)^{\top}$
is a vector of corresponding parameters. We reindex $\boldsymbol{\vartheta}=\left(\boldsymbol{\vartheta}_{1}^{\top},...,\boldsymbol{\vartheta}_{J}^{\top}\right)^{\top}$
for expositional convenience. Letting $\tilde{\boldsymbol{x}}_{\left(h\right),t}=\breve{\boldsymbol{x}}_{t}\otimes\boldsymbol{\varphi}\left(h\right)$,
the model can be expressed as
\begin{eqnarray*}
y_{t+h} & \approx & \boldsymbol{\vartheta}^{\top}\tilde{\boldsymbol{x}}_{\left(h\right),t}+u_{\left(h\right),t+h}.
\end{eqnarray*}
Stacking these equations over the projection dimension yields a representation
à la SUR: 
\[
\boldsymbol{y}_{t}=\tilde{\boldsymbol{X}}_{t}\boldsymbol{\vartheta}+\boldsymbol{u}_{t},
\]
\[
\boldsymbol{y}_{t}=\left(y_{t+h_{1}},...,y_{t+h_{H}}\right)^{\top},\quad\tilde{\boldsymbol{X}}_{t}=\left(\tilde{\boldsymbol{x}}_{\left(h_{1}\right),t},...,\tilde{\boldsymbol{x}}_{\left(h_{H}\right),t}\right)^{\top},\quad\boldsymbol{u}_{t}=\left(u_{\left(h_{1}\right),t+h_{1}},...,u_{\left(h_{H}\right),t+h_{H}}\right)^{\top},
\]
for $t=1,...,T$. Rearranging the above representation delivers

\begin{equation}
\boldsymbol{y}=\tilde{\boldsymbol{X}}\boldsymbol{\vartheta}+\boldsymbol{u},\quad\boldsymbol{u}\sim\mathcal{N}\left(\boldsymbol{0}_{HT},\;\boldsymbol{\Sigma}\otimes\boldsymbol{I}_{T}\right),\label{eq: stacked model w bspline}
\end{equation}
\[
\boldsymbol{y}=\left(\begin{array}{ccc}
\boldsymbol{y}_{\left(h_{1}\right)}^{\top} & \cdots & \boldsymbol{y}_{\left(h_{H}\right)}^{\top}\end{array}\right)^{\top},\quad\tilde{\boldsymbol{X}}=\left(\begin{array}{ccc}
\tilde{\boldsymbol{X}}_{\left(h_{1}\right)}^{\top} & \cdots & \tilde{\boldsymbol{X}}_{\left(h_{H}\right)}^{\top}\end{array}\right)^{\top},\quad\boldsymbol{u}=\left(\boldsymbol{u}_{\left(h_{1}\right)}^{\top},...,\boldsymbol{u}_{\left(h_{H}\right)}^{\top}\right)^{\top},
\]
\[
\boldsymbol{y}_{\left(h\right)}=\left(\begin{array}{ccc}
y_{\left(h\right),1} & \cdots & y_{\left(h\right),T}\end{array}\right)^{\top},\quad\tilde{\boldsymbol{X}}_{\left(h\right)}=\left(\begin{array}{ccc}
\tilde{\boldsymbol{x}}_{\left(h\right),1} & \cdots & \tilde{\boldsymbol{x}}_{\left(h\right),T}\end{array}\right)^{\top},\quad h\in\mathcal{H},
\]
\[
\boldsymbol{u}_{\left(h\right)}=\left(u_{\left(h\right),1},...,u_{\left(h\right),T}\right)^{\top},\quad h\in\mathcal{H}.
\]

We construct a posterior simulator for the model in a similar fashion
as the model without B-spline expansions. Given the same priors for
$\boldsymbol{\tau}$ and $\boldsymbol{\Lambda}$, a prior on $\boldsymbol{\vartheta}$
is constructed as 

\[
p\left(\boldsymbol{\vartheta}|\boldsymbol{\tau},\boldsymbol{\Lambda}\right)\propto\exp\left(-\frac{1}{2}\boldsymbol{\vartheta}^{\top}\boldsymbol{Q}\boldsymbol{\vartheta}\right),
\]
\begin{equation}
\boldsymbol{Q}=\textrm{blkdiag}\left(\tau_{1}\boldsymbol{D}^{\top}\boldsymbol{\Lambda}_{1}\boldsymbol{D},,...,\tau_{J}\boldsymbol{D}^{\top}\boldsymbol{\Lambda}_{J}\boldsymbol{D}\right).\label{eq: prior prec alt}
\end{equation}
The conditional posterior density of $\boldsymbol{\vartheta}$ is
derived as the multivariate normal distribution
\[
\boldsymbol{\vartheta}|-\sim\mathcal{N}\left(\boldsymbol{m},\boldsymbol{P}^{-1}\right),
\]
\begin{eqnarray}
\boldsymbol{m} & = & \boldsymbol{P}^{-1}\tilde{\boldsymbol{X}}^{\top}\left(\boldsymbol{\Sigma}^{-1}\otimes\boldsymbol{I}_{T}\right)\boldsymbol{y},\nonumber \\
\boldsymbol{P} & = & \boldsymbol{Q}+\tilde{\boldsymbol{X}}^{\top}\left(\boldsymbol{\Sigma}^{-1}\otimes\boldsymbol{I}_{T}\right)\tilde{\boldsymbol{X}}.\label{eq: cond_prec_theta_alt}
\end{eqnarray}
As in the model without B-spline expansions, this sampling block presents
a major computational burden. On the one hand, the prior precision
matrix $\boldsymbol{Q}$ can be calculated easily by virtue of its
block diagonal structure (\ref{eq: prior prec alt}) (unless the number
of covariates $J$ is not extremely large). On the other hand, the
quantity $\tilde{\boldsymbol{X}}^{\top}\left(\boldsymbol{\Sigma}^{-1}\otimes\boldsymbol{I}_{T}\right)\tilde{\boldsymbol{X}}$
in (\ref{eq: cond_prec_theta_alt}) involves a high-dimensional matrix
multiplication and cannot be compressed as in (\ref{eq: cond_prec_theta}),
eventually making the posterior simulation more demanding than the
previous case. Sampling distributions of the other parameters are
derived analogously to those of a model without B-spline expansions.

\section{Simulation Study}

We conducted Monte Carlo simulations to investigate the performance
of our proposed approach. We considered six specifications consisting
of the combination of three priors, each with/without B-spline expansions.
As with the N-RP and A-RP priors, we considered a weakly informative
independent standard normal prior, $\boldsymbol{\theta}\sim\mathcal{N}\left(\boldsymbol{0},\;10{}^{4}\boldsymbol{I}\right)$. 

First, we considered a linear data generating processes (DGPs) specified
by the following moving average representation:
\[
y_{t}=\sum_{l=0}^{L}\beta_{\left(l\right)}z_{t-l}+\epsilon_{t},
\]
\[
z_{t}\sim\mathcal{N}\left(0,1\right)\quad,\epsilon_{t}\sim\mathcal{N}\left(0,1\right),
\]
where $y_{t}$ is an endogenous variable, $z_{t}$ is an exogenous
variable, and $\epsilon_{t}$ is the measurement error. A set of parameters
$\beta_{\left(0\right)},...,\beta_{\left(L\right)}$ represents an
impulse response. True parameter values are defined as a convex curve:
\[
\beta_{\left(l\right)}=\frac{l\exp\left(r\left(1-l\right)\right)}{\sum_{l^{\prime}=0}^{L}l^{\prime}\exp\left(r\left(1-l^{\prime}\right)\right)},
\]
\[
r\sim\mathcal{U}\left(0.1,\;1\right),
\]
where $\mathcal{U}\left(0.1,\;1\right)$ denotes a uniform distribution
with support $\left(0.1,\;1\right)$, and $r$ governs where the peak
of the impulse response is located. Covariates are a constant and
four lags of $y_{t}$ and $z_{t}$. We fixed the length of the impulse
response to $L=20$ and the effective sample size to $T=50,100$.
Hyperparameters are $\nu=\nu_{1}=\nu_{2}=0.01$, $\eta=\eta_{1}=\eta_{2}=0.5$,
and $\upsilon=0.01$. We choose the order of the difference matrix
as $r=2$, implying that the sequences of parameters to be inferred
are induced to straight lines. We use the B-spline basis with equidistant
knots ranging from $h_{1}-2$ to $h_{H}-1$ with unitary increments.
We set the degree of the B-spline bases to three. We generate 500
sets of synthetic data. Gibbs sampling obtained 40,000 posterior draws,
after discarding the initial 10,000. Each chain is initialized to
an ordinary least squares estimate. 

We compared the alternative approaches on the basis of four performance
measures: mean squared errors (MSE), coverage probability (Coverage),
lengths of credible intervals (Length), and computational speed (Speed).
MSE is the sum of mean squared errors, 
\[
\text{MSE}=M^{-1}\sum_{m=1}^{M}\sum_{l=0}^{L}\left(\hat{\beta}_{m,\left(l\right)}-\beta_{m,\left(l\right)}^{true}\right)^{2},
\]
where $\hat{\beta}_{m,\left(l\right)}$ denotes a posterior mean estimate
of $\beta_{\left(l\right)}$ in the $m$th experiment, $\beta_{m,\left(l\right)}^{true}$
denotes the corresponding true value, and $M$ is a total number of
experiments. Coverage is the arithmetic mean of the probability that
the true value is within the 90\% credible interval: 
\[
\text{Coverage}=M^{-1}L^{-1}\sum_{m=1}^{M}\sum_{l=0}^{L}\boldsymbol{1}_{\left\{ \hat{\beta}_{m,\left(l\right)}^{5\%}<\beta_{m,\left(l\right)}^{true}\right\} }\times\boldsymbol{1}_{\left\{ \hat{\beta}_{m,\left(l\right)}^{95\%}>\beta_{m,\left(l\right)}^{true}\right\} },
\]
where $\hat{\beta}_{m,\left(l\right)}^{5\%}$ and $\hat{\beta}_{m,\left(l\right)}^{95\%}$
are posterior 5th and 95th percentile estimates of $\beta_{\left(l\right)}$
in the $m$th experiment. Length denotes the arithmetic mean of the
lengths of a 90\% credible interval, 
\[
\text{Length}=M^{-1}L^{-1}\sum_{m=1}^{M}\sum_{l=0}^{L}\left(\hat{\beta}_{m,\left(l\right)}^{95\%}-\hat{\beta}_{m,\left(l\right)}^{5\%}\right).
\]
Speed is the mean computational time of posterior simulations in seconds.\footnote{All programs were written in Matlab 2016a (64 bit) and executed on
an Ubuntu Desktop 16.04 LTS (64 bit), running on Intel Xeon E5-2607
v3 processors (2.6GHz).}

Table 1 reports results of the first experiment. With regard to MSE
and Length, the N-RP and A-RP priors outperform the normal prior,
while the A-RP prior performs slightly worse than the N-RP prior.
Using B-spline expansions reduces MSE and Length but the magnitude
is tiny. Use of the prior and the B-spline does not reduce Coverage.
Speed depends on the prior specification and on whether a B-spline
is used. The difference attributable to the choice of prior is not
notably large, but the use of a B-spline imposes a significant computational
burden. When B-spline expansions are employed, approximately 95\%
of computational time during each MCMC cycle is spent calculating
$\boldsymbol{P}$, in particular, a quantity $\tilde{\boldsymbol{X}}^{\top}\left(\boldsymbol{\Sigma}^{-1}\otimes\boldsymbol{I}_{T}\right)\tilde{\boldsymbol{X}}$.
This bottleneck is a simple matrix-matrix multiplication that is executed
via a built-in mathematical routine of Matlab, so switching to a compiled
language such as Fortran and C/C++ will not totally resolve the problem.
We checked the sensitivity of the simulation results to choice of
the hyperparameters. The results are summarized in the ``Online Appendix''.

That the B-spline function expansions have only a marginal effect
is not surprising, given that response variables can appear as functional
data observed on an equally spaced grid.\footnote{From this point of view, local projections are similar to functional
data models such as \citet{Guo2002,Morris2006}.} Panel (a) in Figure 1 displays B-spline basis functions on a fine
grid (2,401 points) and simulated functional data. This situation
is presumed in a functional data analysis. In a local projection,
however, observation points $\left(h\in\mathcal{H}\right)$ are sparse
and invariant, as demonstrated in panel (b). As is evident from there,
B-spline expansions merely allocate observed information to the fixed
grids, rather than interpolating neighboring information. In our case,
observed information for a single grid point is allocated to neighboring
grid points with weights \{1/6, 3/2, 1/6\}.\footnote{When the degree of the B-spline bases is increased to 5, the weight
set becomes \{1/120, 13/60, 33/60, 13/60, 1/120\}. The added weights
are too small to affect the estimate $\left(1/120\approx0.0083\right)$.} Therefore, using a B-spline indeed smooths estimates of impulse responses,
but its effectiveness is limited. 

We fortified our results by considering nonlinear DGPs characterized
by asymmetric and state-dependent impulse responses, respectively.
The asymmetric DGP is specified by 

\[
y_{t}=\sum_{l=0}^{L}\left(\beta_{\left(l\right)}^{\left[1\right]}z_{t-l}\boldsymbol{1}_{\left\{ z_{t-l}<0\right\} }+\beta_{\left(l\right)}^{\left[2\right]}z_{t-l}\boldsymbol{1}_{\left\{ z_{t-l}\geq0\right\} }\right)+\epsilon_{t},
\]
while the state-dependent DGP is specified by 

\[
y_{t}=\sum_{l=0}^{L}\left(\beta_{\left(l\right)}^{\left[1\right]}z_{t-l}\boldsymbol{1}_{\left\{ y_{t-l}<0\right\} }+\beta_{\left(l\right)}^{\left[2\right]}z_{t-l}\boldsymbol{1}_{\left\{ y_{t-l}\geq0\right\} }\right)+\epsilon_{t},
\]
where $\boldsymbol{1}_{\left\{ \cdot\right\} }$ denotes the indicator
function. For both cases, two sets of parameters are independently
generated in the same way as the linear DGP. We set $T=80,160$. The
other settings are exactly the same as those in the first experiment.
For computational reasons, we did not consider a model with B-spline
expansions. Table 2 presents the results, in which we largely verified
the result of the first experiment. With regard to MSE and Length,
the N-RP and A-RP priors consistently improve accuracy versus the
normal prior. Coverage values obtained using the N-RP and A-RP priors
are slightly smaller than those using the normal prior. 

From the simulation study, we obtained two findings. First, our proposed
approach improves the finite sample performance of local projection,
while such improvements are almost entirely attributable to the roughness
penalty priors and not to the B-spline expansions. Second, despite
its flexibility, the A-RP prior is not superior to the N-RP prior.
In conclusion, a specification with the N-RP prior and no B-spline
expansion is recommendable as a first choice. 

\section{Application}

To demonstrate our model, we applied our approach to an analysis of
monetary policy in the United States. We use monetary policy shocks
compiled by \citet{Coibion2017} which is an update of \citet{Romer2004}.\footnote{The time series of monetary policy shocks is from Yuriy Gorodnichenko's
website (https://eml.berkeley.edu//\textasciitilde{}ygorodni/index.htm). } For the covariates and the response, we considered the following
three macroeconomic variables, downloaded from the Federal Reserve
Economic Data (FRED), maintained by the Federal Reserve Bank of St.
Louis: the industrial production index (FRED mnemonic: INDPRO), the
consumer price index for all urban consumers: all items (CPIAUCSL),
and the effective federal funds rate (FEDFUNDS). We also treated all
three as response variables. We included lags of monetary policy shock
as covariates. All data are monthly and spans from March 1969 to December
2008. The range is limited by the availability of data for monetary
policy shocks. Industrial production and the inflation rate are seasonally
adjusted, and included as annualized month-to-month percentage changes
(log-difference multiplied by 1,200). We included the time trend and
up to four lags of covariates. We choose hyperparameters as in the
previous section. The Gibbs sampler obtained a total of 40,000 posterior
draws after discarding the first 10,000.

Figures 2, 3, and 4 show posterior estimates of the impulse responses
of the macroeconomic variables to monetary policy shocks under different
specifications.\footnote{In the ``Online Appendix'', Figs. A.1, A.2, and A.3 display credible
intervals for all the specifications.} The shaded areas indicate the 90\% credible sets for a preferred
specification using no B-spline and the N-RP prior. For all the response
variables, the roughness penalty priors successfully penalize the
roughness of the impulse response functions. Thus, we obtain economically
plausible, smoothed estimates, and can interpret the shape of the
impulse response easily, recognizing the underlining response. Use
of the B-spline exerts no significant effect on the shape of the impulse
response. 

We then compared the fitness of these estimates based on the deviance
information criterion (DIC) \citep{Spiegelhalter2002} and the Watanabe\textendash Akaike
information criterion (WAIC) \citep{Watanabe2010}. Table 3 reports
on both criteria for different specifications (reported values are
on the deviance scale; the smaller, the better). Specifications including
the roughness penalty priors outperform the normal prior in predictive
accuracy regardless of the fitness measure, while the use of a B-spline
yields only limited improvement. Both B-spline and the roughness penalty
prior enhance fitness, but almost all improvements originate from
the latter.\footnote{Both the DIC and WAIC are asymptotically related to the AIC. Thus,
one might consider evaluating the statistical significance of the
difference in the values of the criteria of two models by applying
a rule of thumb that is originally proposed to Bayes factor. As \citet{Burnham2004}
describe, the AIC can be interpreted as an approximation of the log
marginal likelihood of a model under a \textquotedbl{}savvy\textquotedbl{}
prior that is a function of sample size and the number of model parameters.
According to \citeauthor{Jeffreys1961}'s (1961) rule of thumb, the
statistical significance of the difference between two models is \textquotedbl{}weak\textquotedbl{}
if the difference in the AIC/DIC/WAIC is 0-2, \textquotedbl{}positive\textquotedbl{}
if 2-6, \textquotedbl{}strong\textquotedbl{} if 6-10, or \textquotedbl{}very
strong\textquotedbl{} if >10 (see also \citealp{Raftery1995}). When
this rule of thumb is directly applied to Table 3, one might be able
to interpret the results as follows: the statistical significance
of the differences related to the prior choice is \textquotedbl{}very
strong,\textquotedbl{} and the significance of the differences attributable
to the use of B-splines is \textquotedbl{}weak\textquotedbl{} or \textquotedbl{}positive\textquotedbl{}
when the Normal prior is used while it is \textquotedbl{}very strong\textquotedbl{}
when the N-RP prior is used.} This finding supports our simulation results. When the B-spline is
not used, the posterior simulation takes 26 minutes to generate 50,000
draws; when it is used, the same simulation takes 66 hours. Considering
the higher computational cost, use of a B-spline would not be out
of proportion to the benefit for many applications (Table 3).

\section{Comparison with Existing Approaches}

Recent frequentist approaches to estimate smooth impulse response
are closely related to ours in that their objective functions have
forms similar to the posterior densities we present, i.e., the sum
of a log Gaussian likelihood and a penalty term. From this perspective,
\citet{Barnichon2019} can be seen as a frequentist counterpart to
our approach with both B-spline expansions and roughness penalty priors.
Their objective function is written in our notation as
\[
\hat{\boldsymbol{\vartheta}}=\arg\min\;\left\Vert \boldsymbol{y}-\tilde{\boldsymbol{X}}\boldsymbol{\vartheta}\right\Vert ^{2}+\boldsymbol{\vartheta}^{\top}\left(\tilde{\tau}\boldsymbol{I}_{J}\otimes\boldsymbol{D}^{\top}\boldsymbol{D}\right)\boldsymbol{\vartheta}.
\]
They propose to selecting a (scalar) smoothing parameter $\tilde{\tau}$
using a $k$-fold cross validation. \citeauthor{Barnichon2019}'s
(2019) approach bears only one smoothing parameter, rendering it less
flexible than ours. Figures 5, 6, and 7 plot the posterior estimates
of the (global) smoothing parameters for the real data considered
in Section 4. As evident from these figures, the posterior estimates
of the smoothing parameters are significantly different from covariate
to covariate. Having single smoothing parameter seems implausible
in practice.

\citeauthor{El-Shagi2019}'s (2019) approach can be regarded as a
frequentist version of a model using the N-RP priors and no B-spline
expansion. His estimator is written in our notation as
\[
\hat{\boldsymbol{\theta}}=\arg\min\;\left\Vert \boldsymbol{y}-\tilde{\boldsymbol{X}}\boldsymbol{\theta}\right\Vert ^{2}+\boldsymbol{\theta}^{\top}\left[\boldsymbol{D}^{\top}\boldsymbol{D}\otimes\textrm{diag}\left(\tau_{1},...,\tau_{J}\right)\right]\boldsymbol{\theta}.
\]
This boilds down to a least squares estimator of $\boldsymbol{\theta}$
for an extended model specified by
\[
\left(\begin{array}{c}
\boldsymbol{y}\\
\boldsymbol{0}_{H-r}
\end{array}\right)=\left(\begin{array}{c}
\boldsymbol{I}_{H}\otimes\boldsymbol{X}\\
\boldsymbol{D}\otimes\textrm{diag}\left(\tau_{1}^{1/2},...,\tau_{J}^{1/2}\right)
\end{array}\right)\boldsymbol{\theta}+\left(\begin{array}{c}
\boldsymbol{u}\\
\boldsymbol{u}^{*}
\end{array}\right),
\]
where $\boldsymbol{u}^{*}$ is an $\left(H-r\right)$-dimensional
vector of pseudo residuals generated by the penalty term. Let $\boldsymbol{\Sigma}^{*}$
denote the covariance matrix of $\boldsymbol{u}^{*}$. Given $\boldsymbol{\Sigma}$
and $\boldsymbol{\Sigma}^{*}$, a generalized least squares (GLS)
estimator of $\boldsymbol{\theta}$ is
\begin{eqnarray}
\hat{\boldsymbol{\theta}} & = & \left[\boldsymbol{\Sigma}^{-1}\otimes\left(\boldsymbol{X}^{\top}\boldsymbol{X}\right)+\tilde{\boldsymbol{D}}^{\top}\left(\boldsymbol{\Sigma}^{*}\right)^{-1}\tilde{\boldsymbol{D}}\right]^{-1}\left(\boldsymbol{\Sigma}^{-1}\otimes\boldsymbol{X}^{\top}\right)\boldsymbol{y},\label{eq: GLS esimator}
\end{eqnarray}
\[
\tilde{\boldsymbol{D}}=\boldsymbol{D}\otimes\textrm{diag}\left(\tau_{1}^{1/2},...,\tau_{J}^{1/2}\right).
\]
He chooses $r=2$, restricts $\boldsymbol{\Sigma}$ to be diagonal
and set $\boldsymbol{\Sigma}^{*}$ to a submatrix of $\boldsymbol{\Sigma}$,
that is, 
\[
\boldsymbol{\Sigma}=\textrm{diag}\left(\sigma_{1,1}^{2},...,\sigma_{H,H}^{2}\right),\quad\boldsymbol{\Sigma}^{*}=\textrm{diag}\left(\sigma_{2,2}^{2},...,\sigma_{H-1,H-1}^{2}\right)\otimes\boldsymbol{I}_{J}.
\]
As $\boldsymbol{\Sigma}$ is unknown, the parameters are estimated
through a feasible GLS procedure. First, an ordinary least squares
(OLS) estimate $\hat{\boldsymbol{\theta}}_{OLS}$ is obtained, and
then $\hat{\boldsymbol{\Sigma}}_{OLS}$ is computed using the realized
residuals. Second, using $\hat{\boldsymbol{\Sigma}}_{OLS}$, a first-stage
GLS estimate $\hat{\boldsymbol{\theta}}_{GLS,1}$ is computed as (\ref{eq: GLS esimator})
and compute $\hat{\boldsymbol{\Sigma}}_{GLS,1}$ using the obtained
realized residuals. Lastly, using $\hat{\boldsymbol{\Sigma}}_{GLS,1}$,
a second-stage (final) GLS estimate $\hat{\boldsymbol{\theta}}=\hat{\boldsymbol{\theta}}_{GLS,2}$
is obtained. He chooses $\boldsymbol{\tau}=\left\{ \tau_{1},...,\tau_{J}\right\} $
by minimizing the finite sample corrected Akaike's information criterion
(AICc) \citep{Hurvich1998},
\[
AIC_{c}\left(\boldsymbol{\tau}\right)=-2\log p\left(\mathcal{D}|\hat{\boldsymbol{\theta}}_{GLS,2},\hat{\boldsymbol{\Sigma}}_{GLS,1}\right)+2\delta+\frac{2\delta\left(\delta+1\right)}{T-\delta-1},
\]
 or a variant of the Bayesian information criterion (BICc) analogously
defined as the AICc,
\[
BIC_{c}\left(\boldsymbol{\tau}\right)=-2\log p\left(\mathcal{D}|\hat{\boldsymbol{\theta}}_{GLS,2},\hat{\boldsymbol{\Sigma}}_{GLS,1}\right)+\left(\log T\right)\delta+\frac{2\delta\left(\delta+1\right)}{T-\delta-1}.
\]
$\delta$ denotes the effective loss of degrees of freedom (or pseudo
dimension of the model) which is defined as the trace of a hat (or
projection) matrix $\hat{\boldsymbol{P}}$ with $\hat{\boldsymbol{y}}=\hat{\boldsymbol{P}}\boldsymbol{y}$,
that is, $\delta=\textrm{tr}\left\{ \hat{\boldsymbol{P}}\right\} $
with

\[
\hat{\boldsymbol{P}}=\left[\boldsymbol{\Sigma}^{-1}\otimes\left(\boldsymbol{X}^{\top}\boldsymbol{X}\right)+\left(\boldsymbol{D}^{\top}\left(\boldsymbol{\Sigma}^{*}\right)^{-1}\boldsymbol{D}\right)\otimes\textrm{diag}\left(\tau_{1},...,\tau_{J}\right)\right]^{-1}\left[\boldsymbol{\Sigma}^{-1}\otimes\left(\boldsymbol{X}^{\top}\boldsymbol{X}\right)\right].
\]
In terms of nonparametric regressions, $\delta$ measures the effective
number of zero-th order polynomial bases defined over the projection
horizons $\mathcal{H}$, $\boldsymbol{x}_{t},....,\boldsymbol{x}_{t}$.
This approach can be crudely interpreted as a maximum a posteriori
estimation of a local projection using a (non-adaptive) roughness
penalty prior of $\boldsymbol{\theta}$, a ``prior'' of $\boldsymbol{\tau}$
generated from the AICc or BICc, and a non-informative prior of $\boldsymbol{\Sigma}$.
Figures 8, 9, and 10 represent estimated IRFs of monetary policy shocks
using \citeauthor{El-Shagi2019}'s (2019) approach along with the
default Bayesian estimates. The IRFs obtained by both approaches are
fairly comparable. 

We can identify three advantages of the proposed Bayesian approach
over the frequentist approaches. First, our approach can provide credible
intervals in a consistent and straightforward manner. In contrast,
at this time, the frequentist approaches have no statistically grounded
method to estimate confidence intervals; \citet{Barnichon2019} mention
a heuristic method, while \citet{El-Shagi2019} does not discuss a
method to estimate confidence intervals. 

Second, while frequentist approaches fix smoothing parameters before
inference by cross validation \citep{Barnichon2019} or penalized
optimization \citep{El-Shagi2019}, our approach infers them using
priors, allowing us to systematically consider uncertainty in the
smoothness of an impulse response (and other sequences of coefficients).
The quantitative significance of this conceptual advantage depends
on context. We re-estimated the model with the (global) smoothing
parameters fixed to the posterior medians for the default specifications.
As shown in Figure 11, for the real data in Section 5, there is no
significant difference.

Third, the proposed approach has better finite-sample performance.
We compared \citeauthor{El-Shagi2019}'s (2019) approach to ours through
a simulation study. The simulation setup is the same as that of the
linear IRF in Section 4. \footnote{We minimize the information criteria using the limited-memory Broyden-Fletcher-Goldfarb-Shanno
algorithm with bounds \citep{Byrd1995}, using a Matlab routine \texttt{minConf\_TMP.m}
written by Mark Schmidt.

(https://www.cs.ubc.ca/\textasciitilde{}schmidtm/Software/minConf.html)} We examine specifications with unrestricted and diagonal covariance
matrices for both \citeauthor{El-Shagi2019}'s (2019) and our Bayesian
approaches. A half-t prior is used for the diagonal elements in $\boldsymbol{\Sigma}$,
denoted by $\sigma_{\left(h_{i}\right)}^{2}$, $i=1,...,H$. It is
derived from the HIW prior by setting $H=1$:

\[
\sigma_{\left(h_{i}\right)}^{2}|\phi\sim\mathcal{IG}\left(\frac{\zeta}{2},\;\zeta\phi_{\left(h_{i}\right)}\right),\quad\phi_{\left(h_{i}\right)}\sim\mathcal{G}\left(\frac{1}{2},\upsilon\right),\quad i=1,...,H.
\]
We choose $\upsilon=0.01$ as in Section 3. The result is summarized
in Table 4.\footnote{We also considered Jeffreys prior for $\sigma_{\left(h_{i}\right)}^{2}$
and obtained almost the same result for the half-t prior (thus, it
is not reported).} In line with the simulation study by \citet{El-Shagi2019}, finite
sample performance of the BICc is comparable to or slightly better
than the AICc. The Bayesian approach obtained smaller MSE on average
than the FGLS approach for both covariance specifications. For the
frequentist approach, specifications with diagonal covariances obtained
smaller MSEs than those with unrestricted covariances, whereas for
the Bayesian approach, the situation is the opposite. It is difficult
to identify a specific reason behind this twisted simulation result.
The plug-in estimator of $\boldsymbol{\Sigma}$ employed in the frequentist
approach might not work well for the short time series.\footnote{As $T$ increases (e.g., $T=500$), the relative performance of the
FGLS approach with unrestricted covariance becomes comparable with
that with diagonal covariance (not reported). } The overall winner was the Bayesian approach with unrestricted covariance.
Because residuals in a local projection are strongly correlated by
construction, assuming independence between them is inappropriate. 

\section{Conclusion}

This study developed a fully Bayesian approach to estimate local projections
using roughness penalty priors. It is also considered a specification
involving a B-spline basis function expansion. Monte Carlo experiments
have demonstrated that both B-splines and the roughness penalty priors
improve statistical efficiency, however, almost all the improvements
originate from the latter. Applying our proposed method to an analysis
of monetary policy in the United States shows that the roughness penalty
priors successfully smooth posterior estimates of the impulse response
functions, and can improve the predictive accuracy of local projections. 

This study addresses one of the two disadvantages of local projections,
compared with the standard statistical framework that includes VAR,
namely, that of less statistical efficiency. The other disadvantage
that the exogenous variable is identified ex ante can be resolved
by a two-stage regression approach, as in \citet{Aikman2016}. Constructing
a Bayesian counterpart to this line of approach has not been studied.
In addition, it is potentially beneficial to develop more robust approaches
than ours: for example, a choice of hyperparameters, heteroskedasticity
and autocorrelations in errors, and so on. This study provides a first
step for further developments of Bayesian local projections. 

\paragraph*{Acknowledgement}

The author would like to thank Professor Hideki Konishi for his guidance
and encouragement. The author would also like to thank anonymous referees
as well as participants at the Applied Statistics 2018 for their valuable
suggestions and comments. 

\bibliographystyle{econ}
\bibliography{reference}

\clearpage{}

\begin{table}
\caption{Results of the Monte Carlo simulation: linear IRF}

\medskip{}

\begin{centering}
\begin{tabular}{llccccr}
\hline 
$T$ & Prior & B-spline & MSE & Length & Coverage & Speed\tabularnewline
\hline 
\multirow{6}{*}{50} & Normal &  & 0.542 & 0.976 & 0.997 & 99\tabularnewline
 & Normal & $\surd$ & 0.542 & 0.976 & 0.997 & 1174\tabularnewline
 & N-RP &  & 0.131 & 0.432 & 0.994 & 105\tabularnewline
 & N-RP & $\surd$ & 0.130 & 0.423 & 0.994 & 1164\tabularnewline
 & A-RP &  & 0.150 & 0.468 & 0.994 & 120\tabularnewline
 & A-RP & $\surd$ & 0.151 & 0.459 & 0.993 & 1167\tabularnewline
\hline 
\multirow{6}{*}{100} & Normal &  & 0.243 & 0.599 & 0.995 & 150\tabularnewline
 & Normal & $\surd$ & 0.243 & 0.599 & 0.995 & 4296\tabularnewline
 & N-RP &  & 0.067 & 0.309 & 0.994 & 152\tabularnewline
 & N-RP & $\surd$ & 0.067 & 0.301 & 0.991 & 4291\tabularnewline
 & A-RP &  & 0.074 & 0.331 & 0.995 & 167\tabularnewline
 & A-RP & $\surd$ & 0.074 & 0.323 & 0.993 & 4299\tabularnewline
\hline 
\end{tabular}
\par\end{centering}
\medskip{}

\centering{}%
\begin{minipage}[t]{0.8\columnwidth}%
Note: MSE denotes the mean squared error. Coverage denotes the arithmetic
mean of the probability that the true value is within a 90\% credible
interval. Length denotes the length of the 90\% credible interval.
Speed denotes computational time in seconds.%
\end{minipage}
\end{table}

\clearpage{}

\begin{table}
\caption{Results of the Monte Carlo simulation: nonlinear IRF}

\medskip{}

\begin{centering}
\begin{tabular}{cllccc}
\hline 
$T$ & DGP & Prior & MSE & Length & Coverage\tabularnewline
\hline 
\multirow{6}{*}{80} & \multirow{3}{*}{Asymmetric} & Normal & 2.489 & 1.439 & 0.997\tabularnewline
 &  & N-RP & 0.564 & 0.618 & 0.994\tabularnewline
 &  & A-RP & 0.643 & 0.670 & 0.994\tabularnewline
\cline{2-6} 
 & \multirow{3}{*}{State-dependent} & Normal & 1.526 & 1.073 & 0.996\tabularnewline
 &  & N-RP & 0.453 & 0.500 & 0.985\tabularnewline
 &  & A-RP & 0.504 & 0.541 & 0.987\tabularnewline
\hline 
\multirow{6}{*}{160} & \multirow{3}{*}{Asymmetric} & Normal & 1.211 & 0.911 & 0.993\tabularnewline
 &  & N-RP & 0.339 & 0.445 & 0.988\tabularnewline
 &  & A-RP & 0.373 & 0.477 & 0.989\tabularnewline
\cline{2-6} 
 & \multirow{3}{*}{State-dependent} & Normal & 0.648 & 0.666 & 0.993\tabularnewline
 &  & N-RP & 0.216 & 0.355 & 0.986\tabularnewline
 &  & A-RP & 0.234 & 0.380 & 0.989\tabularnewline
\hline 
\end{tabular}
\par\end{centering}
\medskip{}

\centering{}%
\begin{minipage}[t]{0.8\columnwidth}%
Note: MSE denotes the mean squared error. Coverage denotes the arithmetic
mean of the probability that the true value is within a 90\% credible
interval. Length denotes the length of the 90\% credible interval.%
\end{minipage}
\end{table}

\clearpage{}

\begin{table}
\caption{Comparison of fitness}

\medskip{}

\begin{centering}
\begin{tabular}{lr@{\extracolsep{0pt}.}lr@{\extracolsep{0pt}.}lr@{\extracolsep{0pt}.}lr@{\extracolsep{0pt}.}l}
\hline 
(a) Industrial production & \multicolumn{4}{c}{Normal prior} & \multicolumn{4}{c}{N-RP prior}\tabularnewline
 & \multicolumn{2}{c}{DIC} & \multicolumn{2}{c}{WAIC} & \multicolumn{2}{c}{DIC} & \multicolumn{2}{c}{WAIC}\tabularnewline
\hline 
without B-spline & \multicolumn{2}{c}{32,585} & \multicolumn{2}{c}{31,894} & \multicolumn{2}{c}{31,872} & \multicolumn{2}{c}{31,431}\tabularnewline
with B-spline & \multicolumn{2}{c}{32,583} & \multicolumn{2}{c}{31,894} & \multicolumn{2}{c}{31,854} & \multicolumn{2}{c}{31,423}\tabularnewline
\hline 
 & \multicolumn{2}{c}{} & \multicolumn{2}{c}{} & \multicolumn{2}{c}{} & \multicolumn{2}{c}{}\tabularnewline
\hline 
(b) Inflation & \multicolumn{4}{c}{Normal prior} & \multicolumn{4}{c}{N-RP prior}\tabularnewline
 & \multicolumn{2}{c}{DIC} & \multicolumn{2}{c}{WAIC} & \multicolumn{2}{c}{DIC} & \multicolumn{2}{c}{WAIC}\tabularnewline
\hline 
without B-spline & \multicolumn{2}{c}{29,955} & \multicolumn{2}{c}{29,073} & \multicolumn{2}{c}{29,321} & \multicolumn{2}{c}{28,675}\tabularnewline
with B-spline & \multicolumn{2}{c}{29,950} & \multicolumn{2}{c}{29,068} & \multicolumn{2}{c}{29,309} & \multicolumn{2}{c}{28,647}\tabularnewline
\hline 
 & \multicolumn{2}{c}{} & \multicolumn{2}{c}{} & \multicolumn{2}{c}{} & \multicolumn{2}{c}{}\tabularnewline
\hline 
(c) Fed funds rate & \multicolumn{4}{c}{Normal prior} & \multicolumn{4}{c}{N-RP prior}\tabularnewline
 & \multicolumn{2}{c}{DIC} & \multicolumn{2}{c}{WAIC} & \multicolumn{2}{c}{DIC} & \multicolumn{2}{c}{WAIC}\tabularnewline
\hline 
without B-spline & \multicolumn{2}{c}{35,457} & \multicolumn{2}{c}{34,901} & \multicolumn{2}{c}{34,756} & \multicolumn{2}{c}{34,455}\tabularnewline
with B-spline & \multicolumn{2}{c}{35,461} & \multicolumn{2}{c}{34,902} & \multicolumn{2}{c}{34,732} & \multicolumn{2}{c}{34,445}\tabularnewline
\hline 
\end{tabular}
\par\end{centering}
\medskip{}

\centering{}%
\begin{minipage}[t]{0.8\columnwidth}%
Note: Values of the DIC (deviance information criterion) and WAIC
(Wanatabe-Akaike information criterion) under different specifications
are reported. All values are on the deviance scale.%
\end{minipage}
\end{table}

\clearpage{}

\begin{table}
\caption{Results of the Monte Carlo simulation: comparison to the frequentist
approach}

\medskip{}

\begin{centering}
\begin{tabular}{lllcccccc}
\hline 
$T$ & Prior & Penalty/Prior & \multicolumn{2}{c}{MSE} & \multicolumn{2}{c}{Length} & \multicolumn{2}{c}{Coverage}\tabularnewline
\hline 
 &  &  & full & diagonal & full & diagonal & full & diagonal\tabularnewline
\hline 
\multirow{3}{*}{50} & \multirow{2}{*}{FGLS } & \multirow{1}{*}{AICc} & 0.221 & 0.162 & \textendash{} & \textendash{} & \textendash{} & \textendash{}\tabularnewline
 &  & \multirow{1}{*}{BICc} & 0.166 & 0.110 & \textendash{} & \textendash{} & \textendash{} & \textendash{}\tabularnewline
\cline{2-9} 
 & \multirow{1}{*}{Bayes} & \multirow{1}{*}{N-RP} & 0.131 & 0.151 & 0.432 & 0.298 & 0.994 & 0.925\tabularnewline
\hline 
\multirow{3}{*}{100} & \multirow{2}{*}{FGLS } & \multirow{1}{*}{AICc} & 0.100 & 0.091 & \textendash{} & \textendash{} & \textendash{} & \textendash{}\tabularnewline
 &  & \multirow{1}{*}{BICc} & 0.100 & 0.091 & \textendash{} & \textendash{} & \textendash{} & \textendash{}\tabularnewline
\cline{2-9} 
 & Bayes & N-RP & 0.067 & 0.076 & 0.309 & 0.213 & 0.994 & 0.925\tabularnewline
\hline 
\end{tabular}
\par\end{centering}
\medskip{}

\centering{}%
\begin{minipage}[t]{0.8\columnwidth}%
Note: MSE denotes the mean squared error. Speed denotes the computational
time in seconds.%
\end{minipage}
\end{table}

\clearpage{}

\begin{figure}
\caption{B-spline basis}

\begin{raggedright}
\begin{minipage}[t]{0.45\textwidth}%
\includegraphics{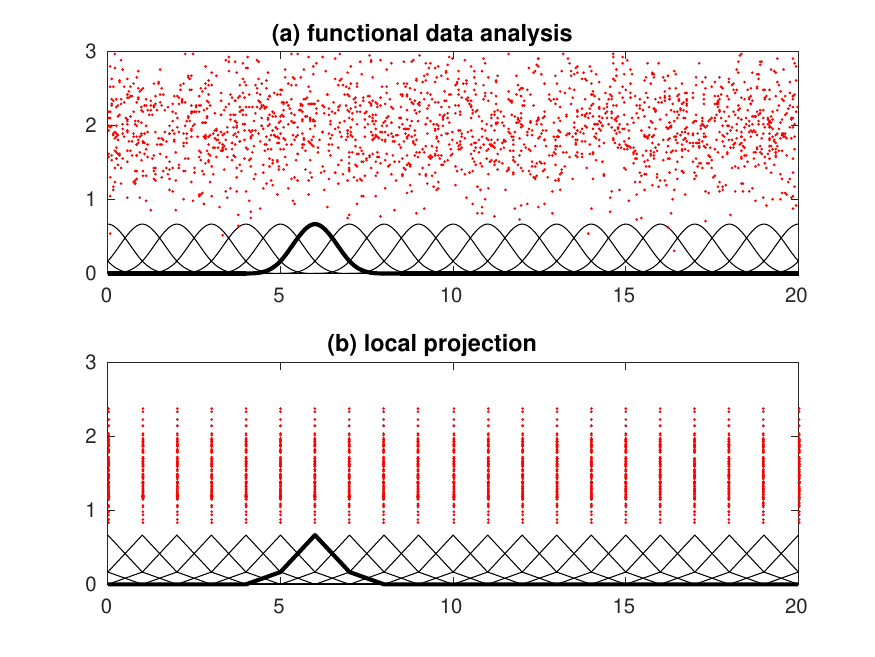}%
\end{minipage}
\par\end{raggedright}
\centering{}\medskip{}
\begin{minipage}[t]{0.8\columnwidth}%
Note: The solid lines show B-spline basis functions used in a functional
data analysis (panel (a)) and local projection (panel (b)), respectively.
Points are simulated observations for each case. Thick lines highlight
the basis functions centered at the sixth knot.%
\end{minipage}
\end{figure}

\clearpage{}

\begin{figure}
\caption{Response of industrial production to monetary policy shocks}

\begin{raggedright}
\begin{minipage}[t]{0.45\textwidth}%
\includegraphics{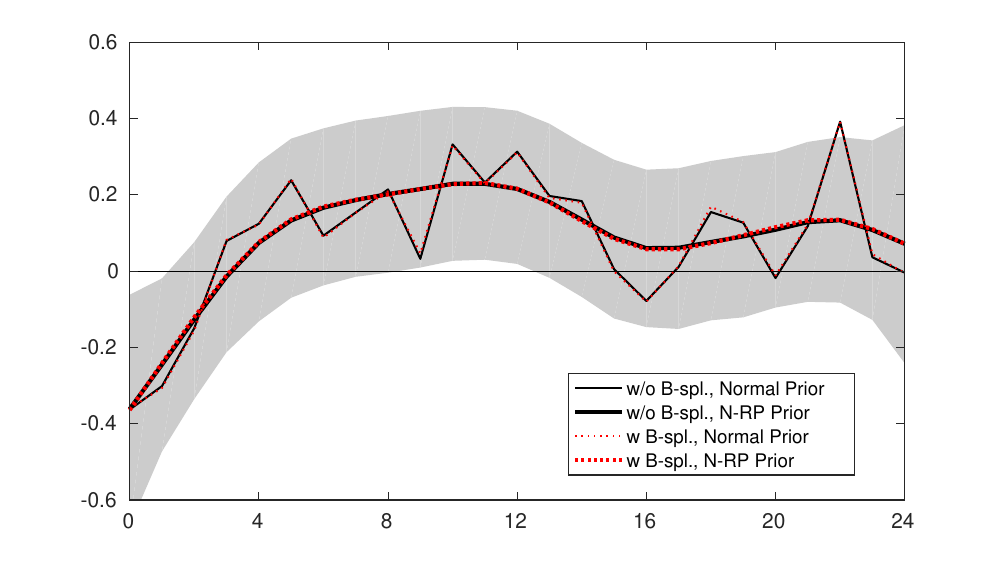}%
\end{minipage}
\par\end{raggedright}
\centering{}\medskip{}
\begin{minipage}[t]{0.8\columnwidth}%
Note: The thin solid line traces the posterior mean for a specification
with no B-spline and the normal prior. The thick solid line traces
the posterior mean for a specification using no B-spline and the N-RP
prior, and the shaded area indicates the corresponding 90\% credible
set. The thin dotted line traces the posterior mean for a specification
with B-splines and the normal prior. The thick dotted line traces
the posterior mean for a specification with B-splines and the N-RP
prior.%
\end{minipage}
\end{figure}

\clearpage{}

\begin{figure}
\caption{Response of inflation to monetary policy shocks}

\begin{raggedright}
\begin{minipage}[t]{0.45\textwidth}%
\includegraphics{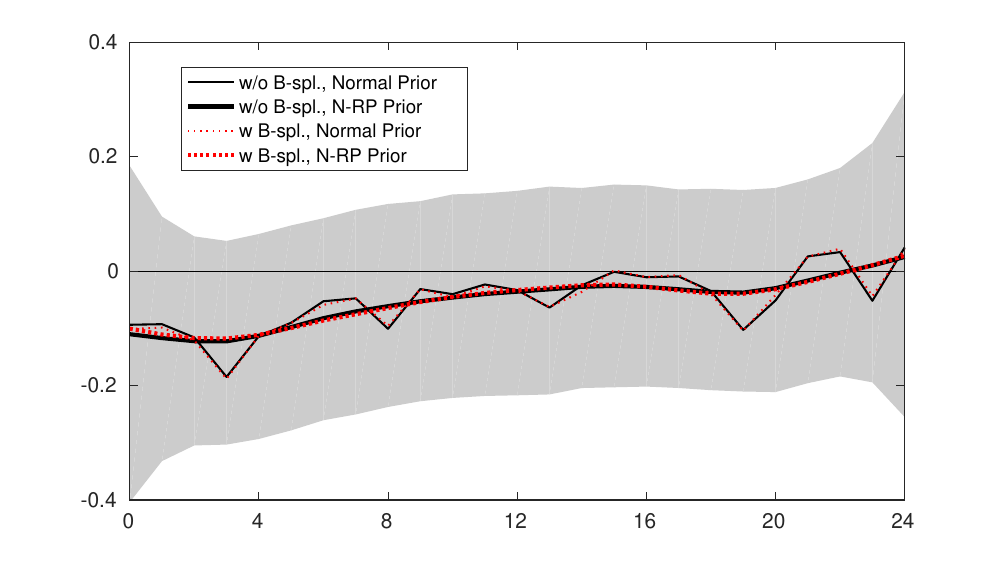}%
\end{minipage}
\par\end{raggedright}
\centering{}\medskip{}
\begin{minipage}[t]{0.8\columnwidth}%
Note: The thin solid line traces the posterior mean for a specification
with no B-spline and the normal prior. The thick solid line traces
the posterior mean for a specification using no B-spline and the N-RP
prior, and the shaded area indicates the corresponding 90\% credible
set. The thin dotted line traces the posterior mean for a specification
with B-splines and the normal prior. The thick dotted line traces
the posterior mean for a specification with B-splines and the N-RP
prior.%
\end{minipage}
\end{figure}

\clearpage{}

\begin{figure}
\caption{Response of fed funds rate to monetary policy shocks}

\begin{raggedright}
\begin{minipage}[t]{0.45\textwidth}%
\includegraphics{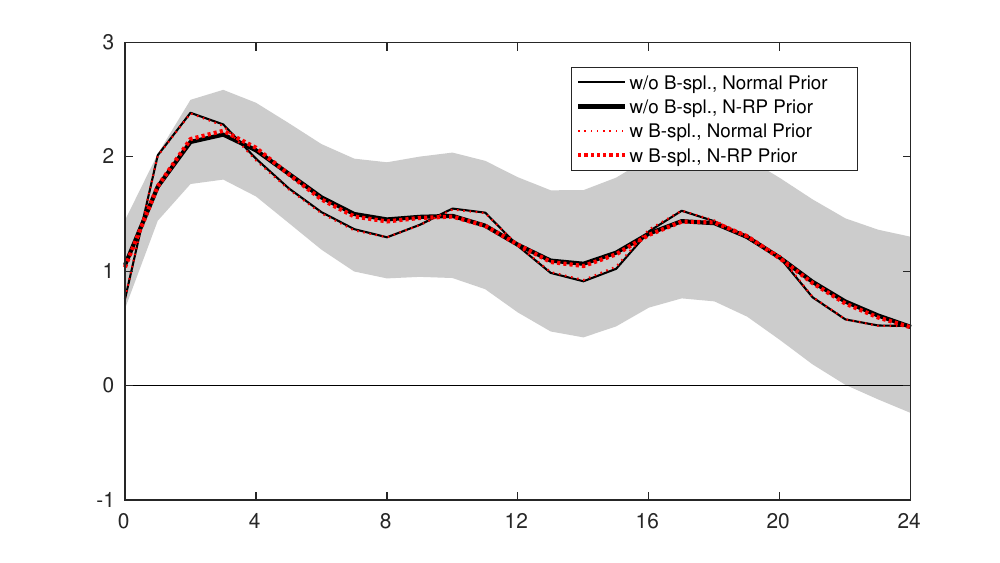}%
\end{minipage}
\par\end{raggedright}
\centering{}\medskip{}
\begin{minipage}[t]{0.8\columnwidth}%
Note: The thin solid line traces the posterior mean for a specification
with no B-spline and the normal prior. The thick solid line traces
the posterior mean for a specification using no B-spline and the N-RP
prior, and the shaded area indicates the corresponding 90\% credible
set. The thin dotted line traces the posterior mean for a specification
with B-splines and the normal prior. The thick dotted line traces
the posterior mean for a specification with B-splines and the N-RP
prior.%
\end{minipage}
\end{figure}

\clearpage{}

\begin{figure}
\caption{Posterior of smoothing parameter: industrial production}

\begin{raggedright}
\begin{minipage}[t]{0.45\textwidth}%
\includegraphics{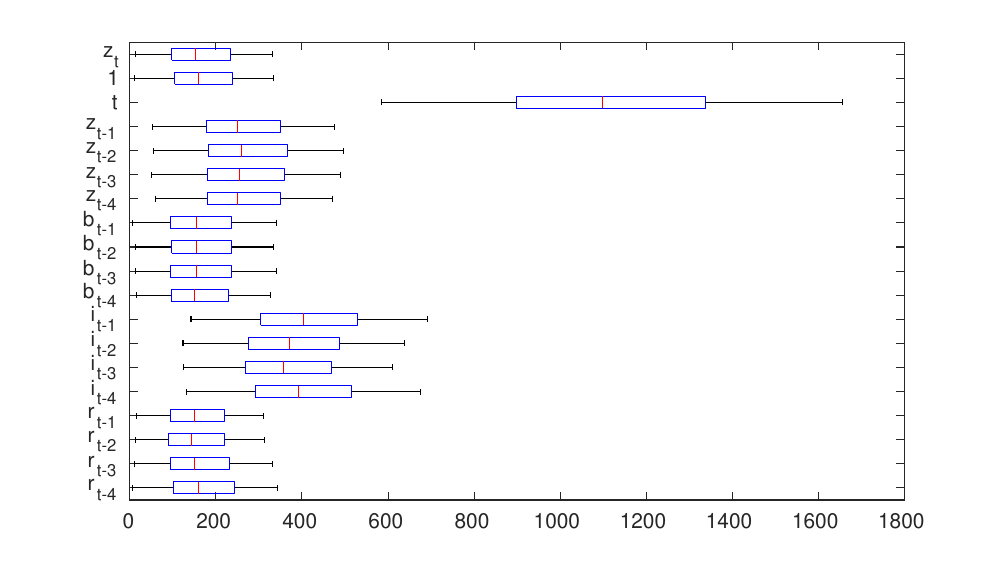}%
\end{minipage}
\par\end{raggedright}
\centering{}\medskip{}
\begin{minipage}[t]{0.8\columnwidth}%
Note: The lines withn the boxes denote the posterior median, the edges
of the boxes denote the 25th and 75th percentiles of the posterior
sample, and the end points of the solid line denote the 5th and 95th
percentiles of the posterior sample. %
\end{minipage}
\end{figure}

\clearpage{}

\begin{figure}
\caption{Posterior of smoothing parameter: inflation}

\begin{raggedright}
\begin{minipage}[t]{0.45\textwidth}%
\includegraphics{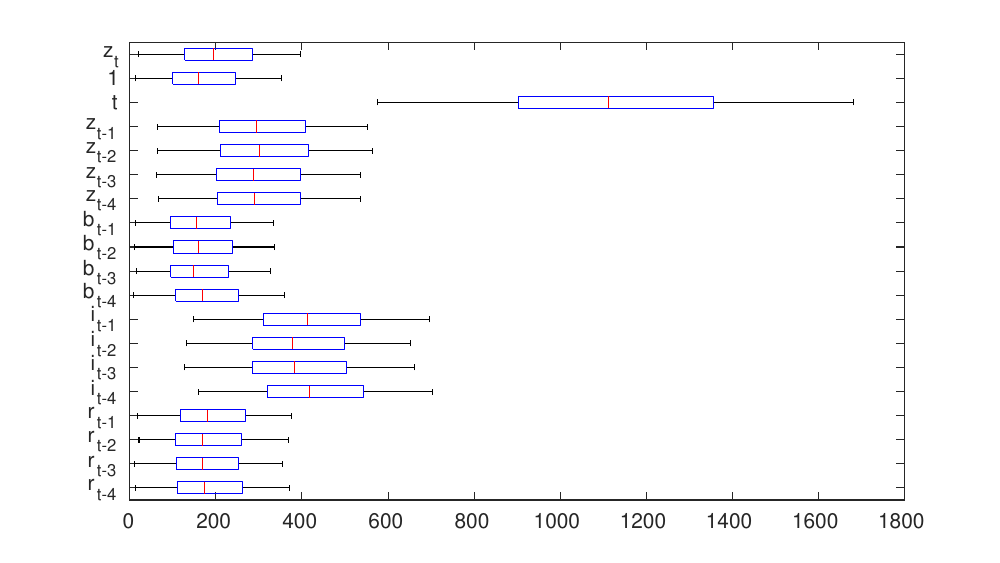}%
\end{minipage}
\par\end{raggedright}
\centering{}\medskip{}
\begin{minipage}[t]{0.8\columnwidth}%
Note: The lines withn the boxes denote the posterior median, the edges
of the boxes denote the 25th and 75th percentiles of the posterior
sample, and the end points of the solid line denote the 5th and 95th
percentiles of the posterior sample. %
\end{minipage}
\end{figure}

\clearpage{}

\begin{figure}
\caption{Posterior of smoothing parameter: Fed funds rate}

\begin{raggedright}
\begin{minipage}[t]{0.45\textwidth}%
\includegraphics{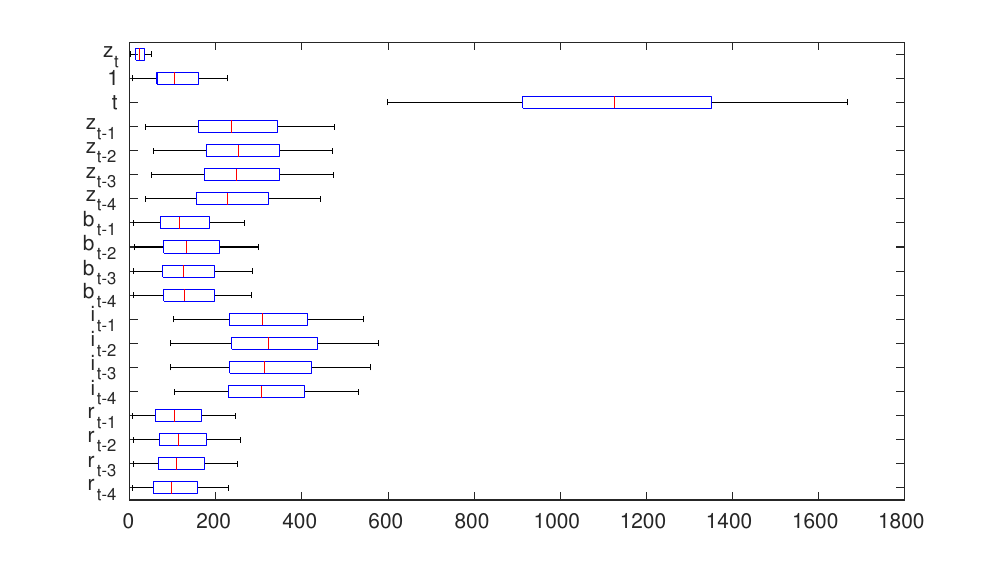}%
\end{minipage}
\par\end{raggedright}
\centering{}\medskip{}
\begin{minipage}[t]{0.8\columnwidth}%
Note: The lines withn the boxes denote the posterior median, the edges
of the boxes denote the 25th and 75th percentiles of the posterior
sample, and the end points of the solid line denote the 5th and 95th
percentiles of the posterior sample. %
\end{minipage}
\end{figure}

\clearpage{}

\begin{figure}
\caption{Response of industrial production to monetary policy shocks: frequentist
and Bayesian approaches}

\begin{raggedright}
\begin{minipage}[t]{0.45\textwidth}%
\includegraphics{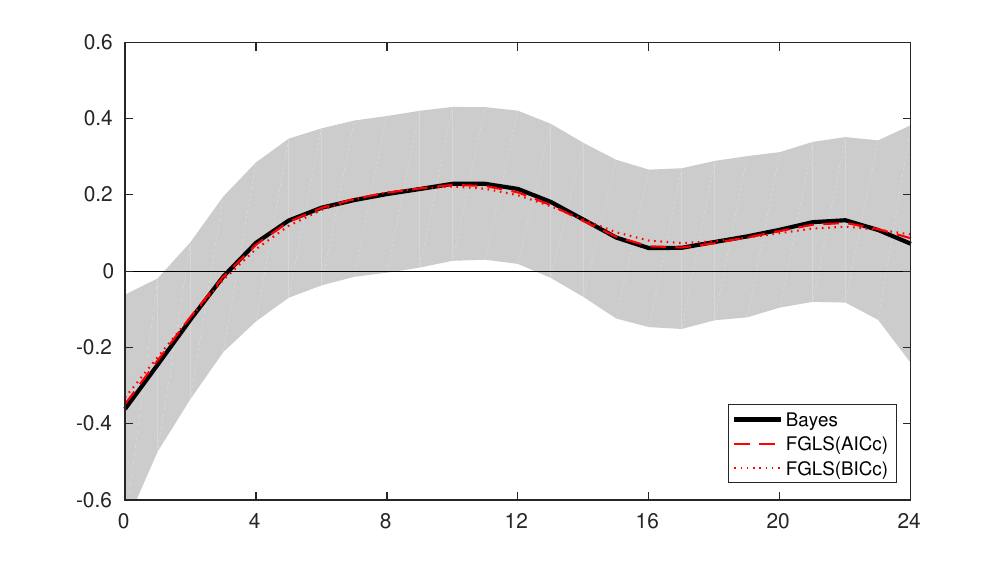}%
\end{minipage}
\par\end{raggedright}
\centering{}\medskip{}
\begin{minipage}[t]{0.8\columnwidth}%
Note: The solid line traces the posterior mean for a Bayesian approach,
and the shaded area indicates the corresponding 90\% credible set.
The dashed line traces an estimate for a frequentist approach with
the AICc. The dotted line traces an estimate for a frequentist approach
with the BICc.%
\end{minipage}
\end{figure}

\clearpage{}

\begin{figure}
\caption{Response of inflation to monetary policy shocks: frequentist and Bayesian
approaches}

\begin{raggedright}
\begin{minipage}[t]{0.45\textwidth}%
\includegraphics{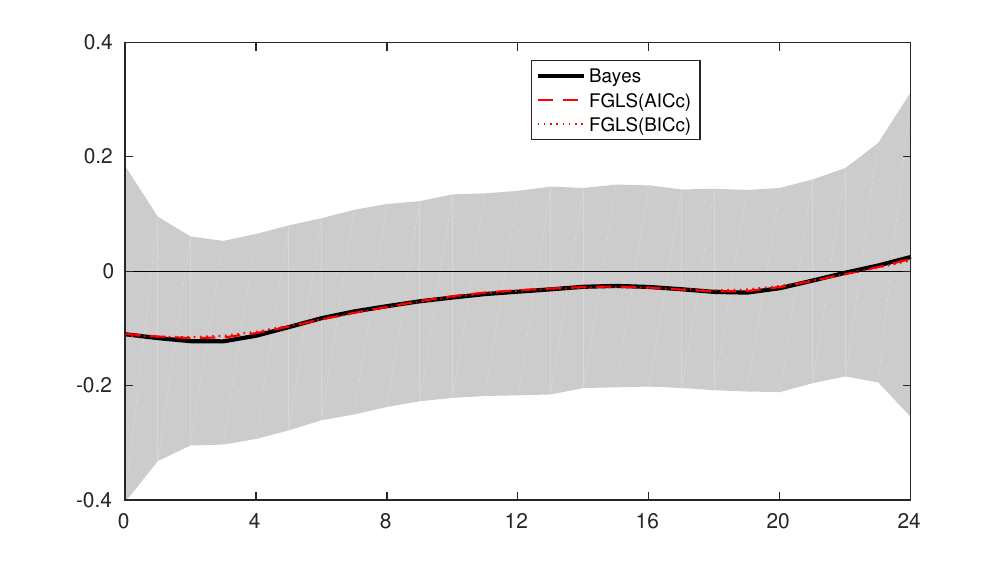}%
\end{minipage}
\par\end{raggedright}
\centering{}\medskip{}
\begin{minipage}[t]{0.8\columnwidth}%
Note: The solid line traces the posterior mean for a Bayesian approach,
and the shaded area indicates the corresponding 90\% credible set.
The dashed line traces an estimate for a frequentist approach with
the AICc. The dotted line traces an estimate for a frequentist approach
with the BICc.%
\end{minipage}
\end{figure}

\clearpage{}

\begin{figure}
\caption{Response of fed funds rate to monetary policy shocks: frequentist
and Bayesian approaches}

\begin{raggedright}
\begin{minipage}[t]{0.45\textwidth}%
\includegraphics{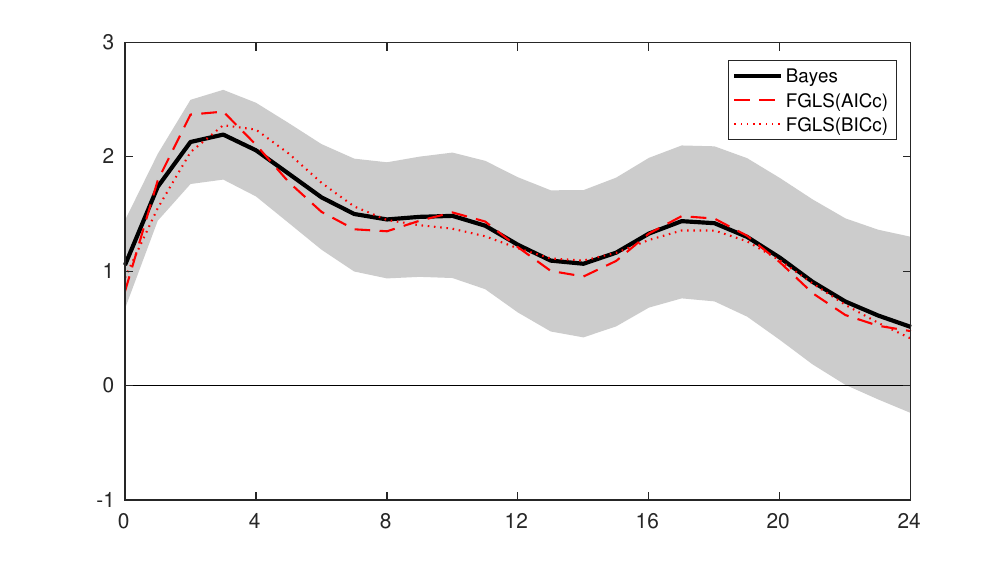}%
\end{minipage}
\par\end{raggedright}
\centering{}\medskip{}
\begin{minipage}[t]{0.8\columnwidth}%
Note: The solid line traces the posterior mean for a Bayesian approach,
and the shaded area indicates the corresponding 90\% credible set.
The dashed line traces an estimate for a frequentist approach with
the AICc. The dotted line traces an estimate for a frequentist approach
with the BICc.%
\end{minipage}
\end{figure}

\clearpage{}

\begin{figure}
\caption{Response to monetary policy shocks: inferred and fixed smoothing parameters}

\begin{raggedright}
\begin{minipage}[t]{0.45\textwidth}%
\includegraphics{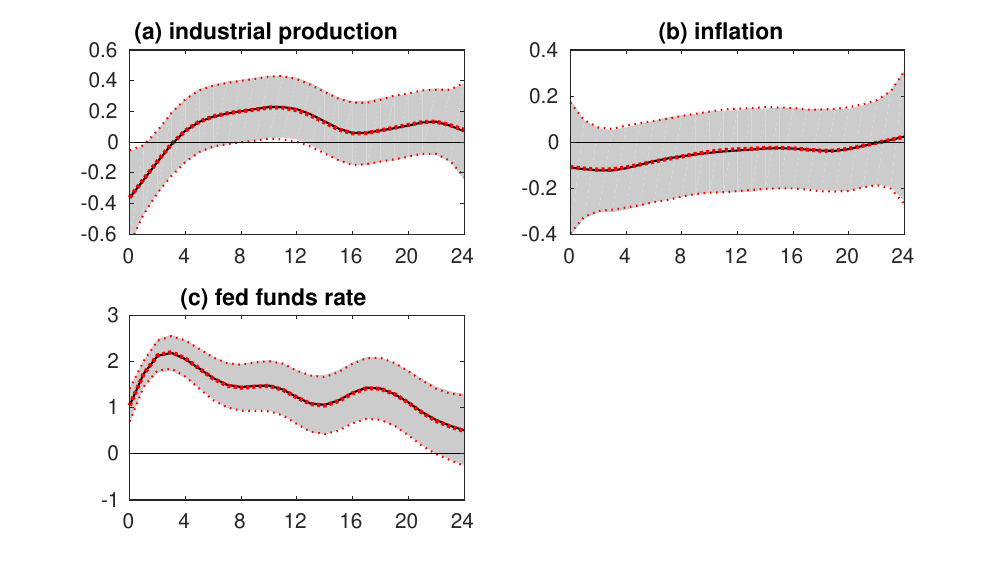}%
\end{minipage}
\par\end{raggedright}
\centering{}\medskip{}
\begin{minipage}[t]{0.8\columnwidth}%
Note: The solid black line traces the posterior mean for a model with
inferred smoothing parameters. The bold dotted line traces the posterior
mean for a model with fixed smoothing parameters. The shaded area
indicates a 90\% credible set for a model with inferred smoothing
parameters. The thin dotted line indicates 90\% credible set for a
model with fixed smoothing parameters.%
\end{minipage}
\end{figure}

\end{document}

\title{Online Appendix to ``Bayesian Inference of Local Projections with
Roughness Penalty Priors''\linebreak{}
(not for publication)}

\author{Masahiro Tanaka}

\date{July 7, 2019}

\maketitle
\setcounter{figure}{0}
\renewcommand{\thefigure}{A.\arabic{figure}}
\setcounter{table}{0} 
\renewcommand{\thetable}{A.\arabic{table}} 

\section*{A.1 Inverse Wishart Prior for $\boldsymbol{\Sigma}$}

This section compare an inverse Wishart prior with the hierarchical
inverse Wishart prior we propose. An inverse Wishart prior is specified
by
\[
\boldsymbol{\Sigma}\sim\mathcal{IW}\left(\boldsymbol{I}_{H},\;\xi+H-1\right),
\]
where $\xi$ is a prefixed hyperparameter. The corresponding conditional
posterior is 
\[
\boldsymbol{\Sigma}|-\sim\mathcal{IW}\left(\boldsymbol{\Xi}+\tilde{\boldsymbol{U}}^{\top}\tilde{\boldsymbol{U}},\;H+\xi+T\right).
\]
This prior is more popular and simpler than the HIW prior. A simulation
environment is the same with the linear case in Section 3 in the main
paper. The simulation result is summarized in Table A.1. Though we
considered $\xi=0,1,2$, the choice of $\xi$ had almost no effect
on the result. Compared to the HIW prior, the inference using the
inverse Wishart prior tended to obtain a shorter length, larger MSE
and less coverage. Based on the simulation result, we prefer the HIW
prior to the inverse Wishart prior. 

\section*{A.2 Sensitivity Check}

We investigated the sensitivity of the inference to the choice of
hyperparameters. A simulation setting is adapted from the linear case
in Section 3 of the main paper. We considered $\nu=0.1,0.01,0.001,0.0001$
for models with the N-RP prior and no B-spline. Table A.2 shows the
results, wherein two items are noteworthy. First, as long as $\nu$
is set between 0.1 and 0.0001, an estimation using the N-RP prior
is more efficient than one using the normal prior. Second, we see
a bias-variance trade-off: as $\nu$ increases (i.e., shrinkage toward
0), an estimator becomes more efficient but less robust. Table A.3
reports the results for $\eta=1,0.5,0.1,0.01$. We can see that as
long as it is chosen within this range, $\eta$ does not significantly
affect the performance. Even when $\eta$ changes, the A-RP prior
connot beat the N-RP prior. Table A.4 includes the results for $\upsilon=0.1,0.01,0.001,0.0001$.
There is almost no difference between the results for the alternative
specifications, which implies that $\upsilon=0.01$ is sufficiently
small for the synthetic data. The results for $r=1,2,3,4$ are shown
in Table A.5. While this experiment indicates that $r=1$ was the
best choice, the approach using the N-RP prior consistently outperformed
that using the normal prior. 

We also conducted a series of sensitivity checks for the real data
application in Section 5 of the main paper using the same alternative
hyperparameter values above. Figures A.1 to A.4 depict the results
for $\nu$, $\eta$, $\upsilon$, and $r$, respectively. We see that
choice of the hyperparameters did not affect the shape of the impulse
response functions.

\clearpage{}

\begin{table}
\caption{Results of the Monte Carlo simulation: inverse Wishart prior for $\boldsymbol{\Sigma}$ }

\medskip{}

\begin{centering}
\begin{tabular}{llr@{\extracolsep{0pt}.}lccc}
\hline 
$T$ & Prior & \multicolumn{2}{c}{$\xi$} & MSE & Length & Coverage\tabularnewline
\hline 
\multirow{4}{*}{50} & HIW & \multicolumn{2}{c}{\textendash{}} & 0.131 & 0.432 & 0.994\tabularnewline
\cline{2-7} 
 & \multirow{3}{*}{IW} & \multicolumn{2}{c}{0} & 0.227 & 0.299 & 0.851\tabularnewline
 &  & \multicolumn{2}{c}{1} & 0.228 & 0.297 & 0.849\tabularnewline
 &  & \multicolumn{2}{c}{2} & 0.229 & 0.296 & 0.842\tabularnewline
\hline 
\multirow{4}{*}{100} & HIW & \multicolumn{2}{c}{\textendash{}} & 0.067 & 0.309 & 0.994\tabularnewline
\cline{2-7} 
 & \multirow{3}{*}{IW} & \multicolumn{2}{c}{0} & 0.090 & 0.217 & 0.906\tabularnewline
 &  & \multicolumn{2}{c}{1} & 0.090 & 0.216 & 0.905\tabularnewline
 &  & \multicolumn{2}{c}{2} & 0.090 & 0.215 & 0.904\tabularnewline
\hline 
\end{tabular}
\par\end{centering}
\medskip{}

\centering{}%
\begin{minipage}[t]{0.8\columnwidth}%
Note: MSE denotes the mean squared error. Coverage denotes the arithmetic
mean of the probability that the true value is within a 90\% credible
interval. Length denotes the length of the 90\% credible interval.%
\end{minipage}
\end{table}

\clearpage{}

\begin{table}
\caption{Results of the Monte Carlo simulation: sensitivity to the choice of
$\nu$ }

\medskip{}

\begin{centering}
\begin{tabular}{llr@{\extracolsep{0pt}.}lccc}
\hline 
$T$ & Prior & \multicolumn{2}{c}{$\nu$} & MSE & Length & Coverage\tabularnewline
\hline 
\multirow{5}{*}{50} & Normal & \multicolumn{2}{c}{\textendash{}} & 0.542 & 0.976 & 0.997\tabularnewline
\cline{2-7} 
 & \multirow{4}{*}{N-RP} & 0&1 & 0.180 & 0.526 & 0.995\tabularnewline
 &  & 0&01 & 0.131 & 0.432 & 0.994\tabularnewline
 &  & 0&001 & 0.099 & 0.367 & 0.993\tabularnewline
 &  & 0&0001 & 0.079 & 0.324 & 0.993\tabularnewline
\hline 
\multirow{5}{*}{100} & Normal & \multicolumn{2}{c}{\textendash{}} & 0.243 & 0.599 & 0.995\tabularnewline
\cline{2-7} 
 & \multirow{4}{*}{N-RP} & 0&1 & 0.090 & 0.378 & 0.995\tabularnewline
 &  & 0&01 & 0.067 & 0.309 & 0.994\tabularnewline
 &  & 0&001 & 0.053 & 0.261 & 0.990\tabularnewline
 &  & 0&0001 & 0.043 & 0.228 & 0.989\tabularnewline
\hline 
\end{tabular}
\par\end{centering}
\medskip{}

\centering{}%
\begin{minipage}[t]{0.8\columnwidth}%
Note: MSE denotes the mean squared error. Coverage denotes the arithmetic
mean of the probability that the true value is within a 90\% credible
interval. Length denotes the length of the 90\% credible interval.%
\end{minipage}
\end{table}

\clearpage{}

\begin{table}
\caption{Results of the Monte Carlo simulation: sensitivity to the choice of
$\eta$ }

\medskip{}

\begin{centering}
\begin{tabular}{llr@{\extracolsep{0pt}.}lccc}
\hline 
$T$ & Prior & \multicolumn{2}{c}{$\eta$} & MSE & Length & Coverage\tabularnewline
\hline 
\multirow{6}{*}{50} & Normal & \multicolumn{2}{c}{\textendash{}} & 0.542 & 0.976 & 0.997\tabularnewline
\cline{2-7} 
 & N-RP & \multicolumn{2}{c}{\textendash{}} & 0.131 & 0.432 & 0.994\tabularnewline
\cline{2-7} 
 & \multirow{4}{*}{A-RP} & 1&0 & 0.143 & 0.454 & 0.994\tabularnewline
 &  & 0&5 & 0.150 & 0.468 & 0.994\tabularnewline
 &  & 0&1 & 0.163 & 0.494 & 0.995\tabularnewline
 &  & 0&01 & 0.157 & 0.481 & 0.994\tabularnewline
\hline 
\multirow{6}{*}{100} & Normal & \multicolumn{2}{c}{\textendash{}} & 0.243 & 0.599 & 0.995\tabularnewline
\cline{2-7} 
 & N-RP & \multicolumn{2}{c}{\textendash{}} & 0.067 & 0.309 & 0.994\tabularnewline
\cline{2-7} 
 & \multirow{4}{*}{A-RP} & 1&0 & 0.072 & 0.323 & 0.994\tabularnewline
 &  & 0&5 & 0.074 & 0.331 & 0.995\tabularnewline
 &  & 0&1 & 0.078 & 0.344 & 0.995\tabularnewline
 &  & 0&01 & 0.074 & 0.332 & 0.994\tabularnewline
\hline 
\end{tabular}
\par\end{centering}
\medskip{}

\centering{}%
\begin{minipage}[t]{0.8\columnwidth}%
Note: MSE denotes the mean squared error. Coverage denotes the arithmetic
mean of the probability that the true value is within a 90\% credible
interval. Length denotes the length of the 90\% credible interval.%
\end{minipage}
\end{table}

\clearpage{}

\begin{table}
\caption{Results of the Monte Carlo simulation: sensitivity to the choice of
$\upsilon$ }

\medskip{}

\begin{centering}
\begin{tabular}{ccr@{\extracolsep{0pt}.}lccc}
\hline 
$T$ & Prior & \multicolumn{2}{c}{$\upsilon$} & MSE & Length & Coverage\tabularnewline
\hline 
\multirow{5}{*}{50} & Normal & \multicolumn{2}{c}{\textendash{}} & 0.542 & 0.976 & 0.997\tabularnewline
\cline{2-7} 
 & \multirow{4}{*}{N-RP} & 0&1 & 0.130 & 0.438 & 0.995\tabularnewline
 &  & 0&01 & 0.131 & 0.432 & 0.994\tabularnewline
 &  & 0&001 & 0.131 & 0.432 & 0.994\tabularnewline
 &  & 0&0001 & 0.131 & 0.432 & 0.994\tabularnewline
\hline 
\multirow{5}{*}{100} & Normal & \multicolumn{2}{c}{\textendash{}} & 0.243 & 0.599 & 0.995\tabularnewline
\cline{2-7} 
 & \multirow{4}{*}{N-RP} & 0&1 & 0.067 & 0.314 & 0.995\tabularnewline
 &  & 0&01 & 0.067 & 0.309 & 0.994\tabularnewline
 &  & 0&001 & 0.067 & 0.309 & 0.994\tabularnewline
 &  & 0&0001 & 0.067 & 0.309 & 0.993\tabularnewline
\hline 
\end{tabular}
\par\end{centering}
\medskip{}

\centering{}%
\begin{minipage}[t]{0.8\columnwidth}%
Note: MSE denotes the mean squared error. Coverage denotes the arithmetic
mean of the probability that the true value is within a 90\% credible
interval. Length denotes the length of the 90\% credible interval.%
\end{minipage}
\end{table}

\clearpage{}

\begin{table}
\caption{Results of the Monte Carlo simulation: sensitivity to choice of $r$ }

\medskip{}

\begin{centering}
\begin{tabular}{llr@{\extracolsep{0pt}.}lccc}
\hline 
$T$ & Prior & \multicolumn{2}{c}{$r$} & MSE & Length & Coverage\tabularnewline
\hline 
\multirow{5}{*}{50} & Normal & \multicolumn{2}{c}{\textendash{}} & 0.542 & 0.976 & 0.997\tabularnewline
\cline{2-7} 
 & \multirow{4}{*}{N-RP} & \multicolumn{2}{c}{1} & 0.077 & 0.381 & 0.998\tabularnewline
 &  & \multicolumn{2}{c}{2} & 0.131 & 0.432 & 0.994\tabularnewline
 &  & \multicolumn{2}{c}{3} & 0.169 & 0.471 & 0.993\tabularnewline
 &  & \multicolumn{2}{c}{4} & 0.197 & 0.503 & 0.991\tabularnewline
\hline 
\multirow{5}{*}{100} & Normal & \multicolumn{2}{c}{\textendash{}} & 0.243 & 0.599 & 0.995\tabularnewline
\cline{2-7} 
 & \multirow{4}{*}{N-RP} & \multicolumn{2}{c}{1} & 0.046 & 0.287 & 0.998\tabularnewline
 &  & \multicolumn{2}{c}{2} & 0.067 & 0.309 & 0.994\tabularnewline
 &  & \multicolumn{2}{c}{3} & 0.081 & 0.328 & 0.990\tabularnewline
 &  & \multicolumn{2}{c}{4} & 0.090 & 0.344 & 0.990\tabularnewline
\hline 
\end{tabular}
\par\end{centering}
\medskip{}

\centering{}%
\begin{minipage}[t]{0.8\columnwidth}%
Note: MSE denotes the mean squared error. Coverage denotes the arithmetic
mean of the probability that the true value is within a 90\% credible
interval. Length denotes the length of the 90\% credible interval.%
\end{minipage}
\end{table}

\clearpage{}

\begin{figure}
\caption{Response of industrial production to monetary policy shocks}

\begin{raggedright}
\begin{minipage}[t]{0.45\textwidth}%
\includegraphics{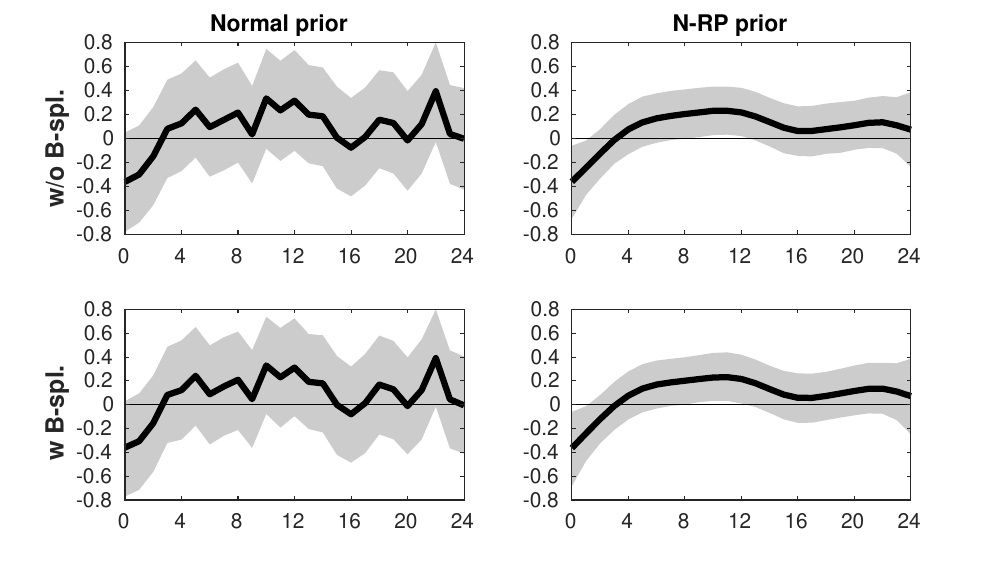}%
\end{minipage}
\par\end{raggedright}
\centering{}\medskip{}
\begin{minipage}[t]{0.8\columnwidth}%
Note: The thick lines trace the posterior mean. The shaded area indicates
the 90\% credible set.%
\end{minipage}
\end{figure}

\clearpage{}

\begin{figure}
\caption{Response of inflation to monetary policy shocks}

\begin{raggedright}
\begin{minipage}[t]{0.45\textwidth}%
\includegraphics{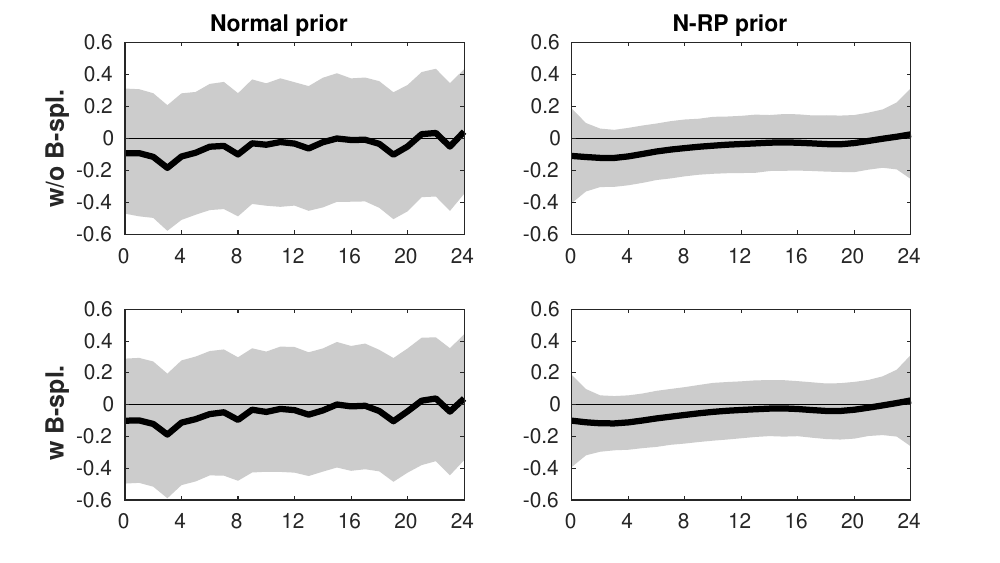}%
\end{minipage}\medskip{}
\par\end{raggedright}
\centering{}%
\begin{minipage}[t]{0.8\columnwidth}%
Note: The thick lines trace the posterior mean. The shaded area indicates
the 90\% credible set.%
\end{minipage}
\end{figure}

\clearpage{}

\begin{figure}
\caption{Response of the Fed funds rate to monetary policy shocks}

\begin{raggedright}
\begin{minipage}[t]{0.45\textwidth}%
\includegraphics{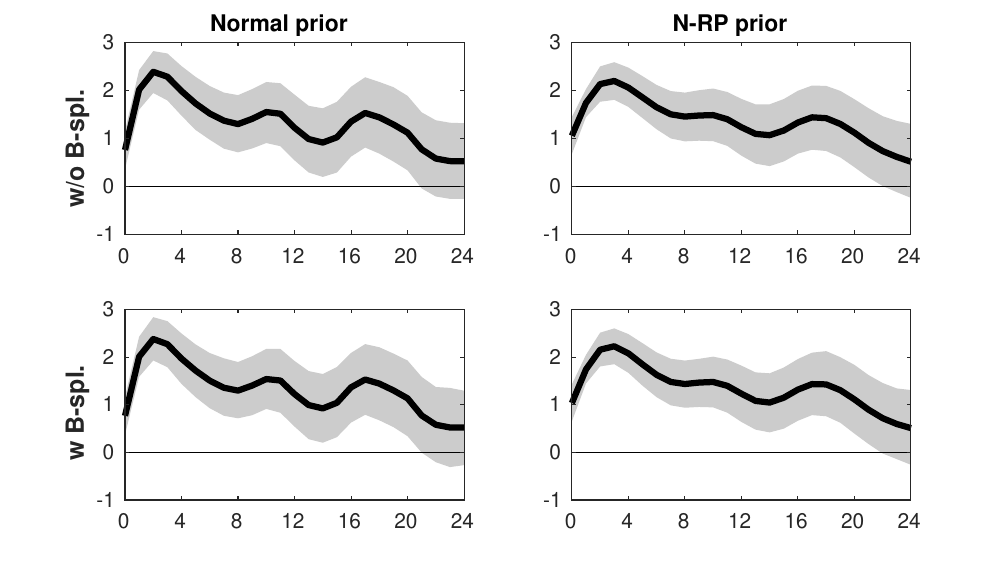}%
\end{minipage}
\par\end{raggedright}
\centering{}\medskip{}
\begin{minipage}[t]{0.8\columnwidth}%
Note: Thick lines trace the posterior mean. The shaded area indicates
the 90\% credible set.%
\end{minipage}
\end{figure}

\clearpage{}

\begin{figure}
\caption{Response to monetary policy shocks: sensitivity to $\nu$}

\begin{raggedright}
\begin{minipage}[t]{0.45\textwidth}%
\includegraphics{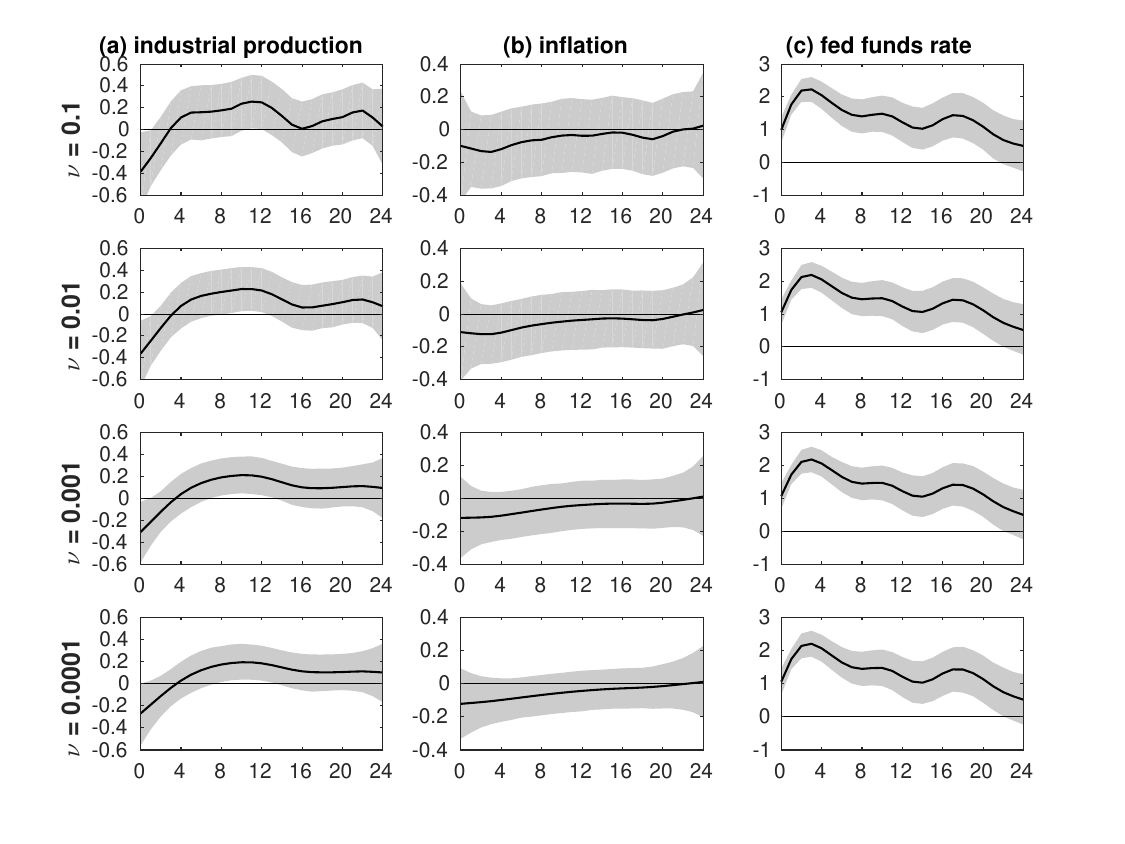}%
\end{minipage}
\par\end{raggedright}
\centering{}\medskip{}
\begin{minipage}[t]{0.8\columnwidth}%
Note: Thick lines trace the posterior mean. Shaded area indicates
90\% credible set.%
\end{minipage}
\end{figure}

\clearpage{}

\begin{figure}
\caption{Response to monetary policy shocks: sensitivity to $\eta$}

\begin{raggedright}
\begin{minipage}[t]{0.45\textwidth}%
\includegraphics{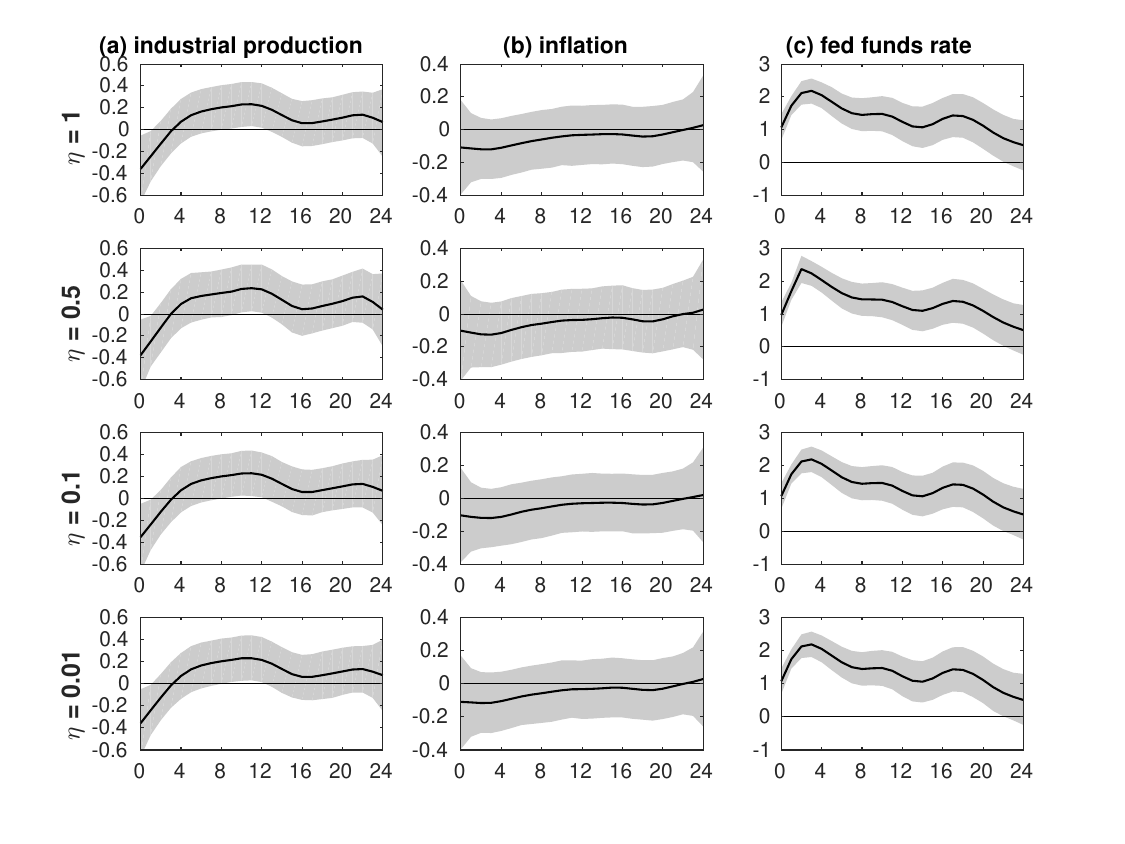}%
\end{minipage}
\par\end{raggedright}
\centering{}\medskip{}
\begin{minipage}[t]{0.8\columnwidth}%
Note: Thick lines trace the posterior mean. Shaded area indicates
90\% credible set.%
\end{minipage}
\end{figure}

\clearpage{}

\begin{figure}
\caption{Response to monetary policy shocks: sensitivity to $\upsilon$}

\begin{raggedright}
\begin{minipage}[t]{0.45\textwidth}%
\includegraphics{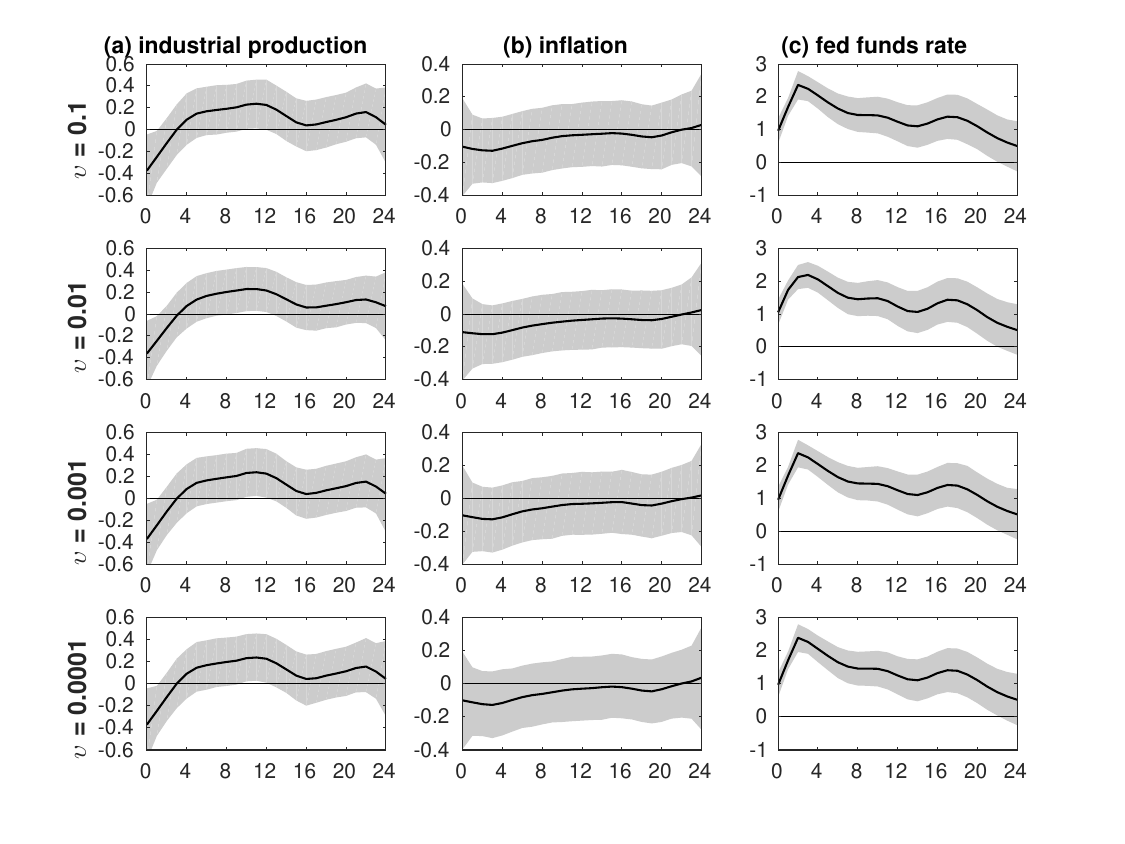}%
\end{minipage}
\par\end{raggedright}
\centering{}\medskip{}
\begin{minipage}[t]{0.8\columnwidth}%
Note: Thick lines trace the posterior mean. Shaded area indicates
90\% credible set.%
\end{minipage}
\end{figure}

\clearpage{}

\begin{figure}
\caption{Response to monetary policy shocks: sensitivity to $r$}

\begin{raggedright}
\begin{minipage}[t]{0.45\textwidth}%
\includegraphics{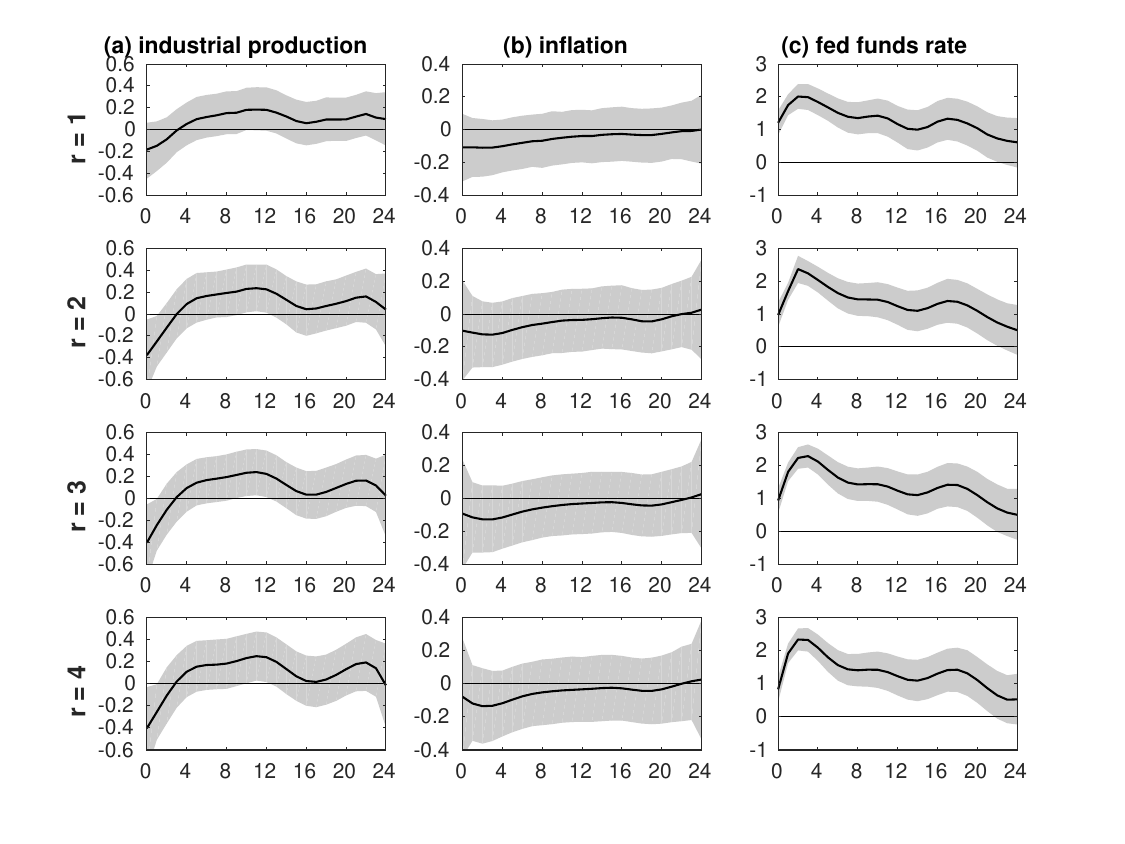}%
\end{minipage}
\par\end{raggedright}
\centering{}\medskip{}
\begin{minipage}[t]{0.8\columnwidth}%
Note: Thick lines trace the posterior mean. Shaded area indicates
90\% credible set.%
\end{minipage}
\end{figure}